\documentclass[aps,epsf,nofootinbib,notitlepage,showpacs,10pt, unsortedaddress, superscriptaddress, twocolumn]{revtex4-1}

\usepackage[svgnames, table]{xcolor}

\usepackage{color}
\usepackage{float}
\usepackage{graphicx}
\usepackage{subcaption}
\usepackage{amsmath}
\usepackage{amsthm}
\usepackage{mathtools}

\theoremstyle{definition}

\usepackage{bm}
\usepackage{hyperref}
\usepackage[capitalise]{cleveref}

\usepackage{float}
\usepackage{multirow}
\restylefloat{table}
\usepackage{algorithm}
\usepackage{algpseudocode}

\newcommand{\bra}[1]{{\left\langle #1 \right|}}
\newcommand{\ket}[1]{{\left| #1 \right\rangle}}

\usepackage{quantikz}
\theoremstyle{definition}
\newtheorem{thm}{Theorem}[section]

\newtheorem{proposition}[thm]{Proposition}
\newtheorem{corollary}[thm]{Corollary}

\definecolor{AW}{HTML}{286D8C}

\usepackage{tikz} 
\usetikzlibrary{matrix,positioning}
\usepackage{amsfonts}
\usepackage{amssymb}

\definecolor{rpLightPurple}{HTML}{232136}
\definecolor{rpDarkPurple}{HTML}{1F1D2E}
\definecolor{rpWhite}{HTML}{e0def4}
\definecolor{rpPink}{HTML}{c4a7e7}
\definecolor{rpBlue}{HTML}{3e8fb0}
\definecolor{rpLightBlue}{HTML}{9ccfd8}
\definecolor{rpYellow}{HTML}{f6c177}
\definecolor{rpRed}{HTML}{eb6f92}
\definecolor{rpGreen}{HTML}{31748f}
\definecolor{rpBlack}{HTML}{26233a}
\definecolor{rpOtherBlack}{HTML}{6e6a86}
\definecolor{rpMagenta}{HTML}{c4a7e7}
\definecolor{lavender}{HTML}{7C7294}
\definecolor{amathyst}{HTML}{9063CD}

\hypersetup{
   colorlinks = true, urlcolor=blue
}
\usepackage[normalem]{ulem}
\usepackage{lipsum}

\graphicspath{{figures/}}
\usepackage{tikz-network}
\usetikzlibrary{fit, shapes.arrows}
\usetikzlibrary{shapes.geometric, arrows}
\usetikzlibrary{positioning,chains,fit,shapes,calc}
\usetikzlibrary{arrows.meta}

\tikzstyle{boxblue} = [rectangle, rounded corners, minimum width=3cm, minimum height=1cm,text centered, draw=black, fill=blue!30]
\tikzstyle{boxred} = [rectangle, rounded corners, minimum width=3cm, minimum height=1cm,text centered, draw=black, fill=red!30]
\tikzstyle{boxgreen} = [rectangle, rounded corners, minimum width=3cm, minimum height=1cm,text centered, draw=black, fill=green!30]
\tikzstyle{io} = [trapezium, trapezium left angle=70, trapezium right angle=110, minimum width=3cm, minimum height=1cm, text centered, draw=black, fill=orange!30]
\tikzstyle{process} = [rectangle, minimum width=3cm, minimum height=1cm, text centered, draw=black, fill=orange!30]
\tikzstyle{decision} = [diamond, minimum width=3cm, minimum height=1cm, text centered, draw=black, fill=green!30]
\tikzstyle{arrow} = [thick,->,>=stealth]

\tikzstyle{decision} = [diamond, draw, fill=blue!20, 
    text width=4.5em, text badly centered, node distance=3cm, inner sep=0pt]
\tikzstyle{block} = [rectangle, draw, fill=blue!20, 
    text width=5em, text centered, rounded corners, minimum height=4em]
\tikzstyle{line} = [draw, -latex']
\tikzstyle{cloud} = [draw, ellipse,fill=red!20, node distance=3cm,
    minimum height=2em]

\usepackage{float}

\usepackage{array, booktabs, makecell, multirow}

\usepackage{amsmath,calc}

\definecolor{graphPurple}{RGB}{185,75,185}

\usepackage{adjustbox}

\begin{document}
\markboth{Wilkie et al.}
{PC-QAOA}

\title{Partitioned-Constraint QAOA (PC-QAOA): Structural State Preparation and Penalty Enforcement for Quantum Optimization}

\author{Anthony Wilkie}
\affiliation{
	Department of Industrial and Systems Engineering\ University of Tennessee at Knoxville \ Knoxville, TN 37996}
    
\author{Alexander DeLise}
\affiliation{
	Department of Mathematics, Department of Scientific Computing \ Florida State University \ Tallahassee, FL 32306}
    
\author{Andrew Del Real}
\affiliation{
	Department of Mathematics, Department of Computer Science\ Carthage College \ Kenosha, WI 53140}

\author{Rebekah Herrman}
\affiliation{
	Department of Industrial and Systems Engineering\ University of Tennessee at Knoxville \ Knoxville, TN 37996}
    
\author{James Ostrowski}\thanks{corresponding author}
\email{jostrows@utk.edu}
\affiliation{
	Department of Industrial and Systems Engineering\ University of Tennessee at Knoxville \ Knoxville, TN 37996}


\begin{abstract}

Constrained combinatorial optimization remains challenging for quantum algorithms because feasibility must be explicitly enforced, typically through penalty terms or problem-specific mixers. We introduce Partitioned-Constraint QAOA (PC-QAOA), which partitions constraints into those enforced structurally and those enforced energetically. Structural constraints are handled via feasible-state preparation and a Grover mixer that preserves feasibility, while the remaining constraints are enforced through penalties. We show that constraints with disjoint support can be prepared in parallel with little error accumulation. We identify broad classes of constraints (including cardinality, assignment, and flow conservation) that admit efficient structural enforcement, and introduce a variational gadget construction that extends this approach to arbitrary low-support constraints. Across 413 completed instances spanning multiple constraint families, PC-QAOA substantially improves feasibility and solution quality at shallow depth relative to penalty-based QAOA, demonstrating the value of partial structural enforcement.

\end{abstract}

\maketitle

\section{Introduction}\label{sec: Introduction}

Quantum optimization algorithms, and in particular the Quantum Approximate Optimization Algorithm (QAOA)~\cite{farhi2014quantum}, have shown promise for solving quadratic unconstrained binary optimization (QUBO) problems~\cite{shaydulin2024evidence, montanez2024towards, dupont2023quantum, cain2023quantum}.
Most combinatorial optimization problems of practical interest, however, are \emph{constrained}.
The dominant strategy used in the literature converts constrained problems to unconstrained problems by modifying the objective function to penalize infeasibility 
~\cite{Hadfield_2019, egger2021warm, karimi2019practical}. This penalization can require auxiliary variables and increase the density of the resulting QUBO.

Alternative approaches encode constraints directly into the mixing Hamiltonians to confine the search to the feasible subspace~\cite{Hadfield_2019, bartschi2020grover, goldstein2024convergence}, use oracle-assisted quantum walks~\cite{marsh2019quantum}, or employ penalty-free QAOA formulations~\cite{bucher2025if, angara2025art}.
These methods eliminate penalty tuning and improve feasibility, but typically require problem-specific mixer circuits whose initial eigenstates can be difficult to prepare~\cite{he2023alignment, tate2022bridgingclassicalquantumsdp}, limiting applicability to structured constraint families.
Machine learning and hybrid classical approaches have also been applied to constrained combinatorial optimization~\cite{cappart2021combining, guo2019solving, heydaribeni2024distributed}. 

This work addresses both limitations by introducing a \emph{hybrid} constraint-handling strategy.
We propose \emph{PC-QAOA} (Partitioned-Constraint QAOA), a framework that partitions constraints into two classes: those that are enforced structurally and those that are enforced via energetic penalties.
Constraints in the structural subset are incorporated directly into the initial state and mixing Hamiltonian by preparing (exact or approximate) superpositions over their feasible assignments, thereby restricting the search to a reduced subspace.
The remaining constraints are enforced through standard penalty terms in the cost Hamiltonian.
This approach interpolates between fully penalty-based QAOA and fully constrained mixer methods, enabling a tradeoff between circuit complexity and search space reduction.

To support this framework, we leverage both exact and variational constructions for feasible-state preparation.
For several constraint families, including cardinality and assignment constraints, the required feasible-state superpositions can be prepared exactly using known circuits (e.g., Dicke-state preparation), enabling efficient and noise-resilient structural enforcement without additional training.

For more general small-support constraints where exact constructions are unavailable or impractical, we introduce \emph{Variational Constraint Gadgets} (VCGs).
These gadgets are trained offline to approximate equal superpositions over the feasible assignments of individual constraints and can be reused across problem instances.
VCGs therefore provide a flexible fallback mechanism for structural enforcement, extending the applicability of PC-QAOA beyond the constraint families mentioned above. 

We evaluate PC-QAOA on a range of constrained combinatorial optimization problems and compare it to standard penalty-based formulations.
Our results show that partial structural enforcement significantly improves the probability of measuring both  a feasible solution and an optimal solution. 
These findings suggest that selectively encoding constraint structure into the initial quantum state and dynamics is a promising direction for scalable constrained quantum optimization.






The remainder of the paper is organized as follows.
Section~\ref{sec: Background} reviews combinatorial optimization problems (COPs) and QAOA.
Section~\ref{sec: methods} surveys existing approaches for encoding constrained COPs into quantum algorithms.
Section~\ref{sec: vcg} introduces VCGs and feasible-state preparation methods.
Section~\ref{sec: hybrid} presents the PC-QAOA framework.
Section~\ref{sec: Results} provides empirical evaluation comparing PC-QAOA to penalty-based QAOA on a large set of randomly generated problem instances,
and Section~\ref{sec: Discussion} concludes with directions for future work.

\section{Background}\label{sec: Background}
\subsection{Combinatorial optimization}\label{sec:cop}
Most optimization problems of interest are naturally thought of as constrained optimization problems (COPs),  where one seeks to optimize a discrete objective subject to a collection of constraints.
In this work, we focus on binary COPs of the form
\begin{align}
    \min_{x \in \{0,1\}^n}\quad & f(x) = x^\top Q x \label{eq:cop_obj} \\
    \text{s.t.}\quad & c_k(x) \le b_k, \quad \forall (c_k, b_k) \in \mathcal{C}, \label{eq:cop_con}
\end{align}
where $Q \in \mathbb{R}^{n \times n}$ is an upper-triangular matrix defining a quadratic objective function, $\mathcal{C}$ is a collection of pairs $(c_k, b_k)$ with $c_k:\mathbb{R}^n \to \mathbb{R}$, and $b_k \in \mathbb{R}$ defining the constraint $c_k(x) \leq b_k$.
This formulation encompasses a broad class of NP-hard problems, such as  knapsack, portfolio optimization, and scheduling problems.

\paragraph{Constraint support.}
For a constraint function $c$ of the form in Eqn.~\eqref{eq:cop_con}, we define its \emph{support} as the set
\begin{equation*}
\mathrm{supp}(c) = \{\, i \in \{1,\dots,n\} \mid c(x) \text{ depends on } x_i \,\}.
\end{equation*}
When $c$ is differentiable, this coincides with the set of indices for which $\partial c(x)/\partial x_i$ is not identically zero.
The support size determines the locality of the constraint and directly impacts the complexity of both classical reformulations and quantum circuit implementations.

\paragraph{Penalty-based reformulation with logarithmic slack.}
For each constraint $c_k(x) \le b_k$~\cite{karimi2019practical}, we introduce a logarithmic (binary) slack variable \[ s_k = \sum_{j=0}^{m_k-1} 2^j y_{kj}, \quad y_{kj} \in \{0,1\}, \] where $m_k = \lceil \log_2 (b_k - c_k^{\min} + 1) \rceil$ and $c_k^{\min}$ is a known lower bound on $c_k(x)$. 
The inequality constraint is then rewritten as the equality \[ c_k(x) + s_k = b_k.
\]

The constrained problem in Eqns.~\eqref{eq:cop_obj}–\eqref{eq:cop_con} can thus be reformulated as the unconstrained optimization problem
\begin{equation*}
    \label{eq:penalty_form}
    \min_{x,s} \; f(x)
    + \sum_{k=1}^{|\mathcal{C}|}
    \delta_k \bigl(c_k(x) + s_k - b_k\bigr)^2,
\end{equation*}
where $\delta_k > 0$ are penalty coefficients that weight the importance of satisfying each constraint. Another common approach is to pick a single penalization constant $\delta$ and multiply it with the entire sum instead of picking custom $\delta_k$ for each constraint. 

\subsection{QAOA}\label{sec:qaoa}

Many COPs over binary variables can be expressed as quadratic unconstrained binary optimization (QUBO) problems.
To solve such problems using quantum algorithms, the QUBO objective is first mapped to a cost Hamiltonian acting on qubits, for which a convenient representation is provided by an Ising Hamiltonian of the form \cite{lucas2014ising}
\begin{equation}\label{eq:ising}
    H_f = - \sum_{\langle i, j \rangle} J_{ij} Z_i Z_j - \sum_{j} h_j Z_j,
    \qquad J_{ij}, h_j \in \mathbb{R},
\end{equation}
where $Z_i = I^{\otimes i-1} \otimes Z \otimes I^{\otimes n - i -1}$ denotes the Pauli-$Z$ operator acting on qubit $i$ and $Z$ is defined as 
\[
Z =\begin{bmatrix}
        1 & 0 \\
        0 & -1
    \end{bmatrix}.
\]
This cost Hamiltonian is constructed by replacing each variable $x_i$ in the COP's objective function using the mapping
\begin{equation*}
    x_i \;\longrightarrow\; \frac{I - Z_i}{2}.
\end{equation*}
The Hamiltonian $H_f$ has eigenvalues equal to $f(x)$ for each bitstring $x$.
Under this substitution, minimizing the QUBO objective is equivalent to finding the ground state of the cost Hamiltonian $H_f$. 

QAOA is a variational algorithm designed to approximately solve COPs using the Hamiltonian in Eqn.~\eqref{eq:ising} by alternating between unitary evolutions generated by the cost Hamiltonian $H_f$ and a mixing Hamiltonian $H_B$, which play complementary roles in the algorithm.

The cost unitary \[ U_f(\gamma) = e^{-i \gamma H_f} \] applies phases to computational basis states (states which represent classical bitstrings) proportional to their objective values.
As a result, basis states corresponding to lower-cost solutions accumulate different phases than higher-cost states.

The mixing unitary \[ U_B(\beta) = e^{-i \beta H_B} \] is generated by a mixing Hamiltonian $H_B$ that induces transitions between computational basis states via amplitude mixing.
A common choice is the $X$-mixer
\begin{equation*}
    H_B = \sum_{j} X_j,
\end{equation*}
where $X_j = I^{\otimes j-1} \otimes X \otimes I^{\otimes n - j -1}$ denotes the Pauli-$X$ operator acting on qubit $j$ and $X$ is defined as 
\[
X =
    \begin{bmatrix}
        0 & 1 \\
        1 & 0
    \end{bmatrix}.
\]
  With this choice, $U_B(\beta) = \prod_j e^{-i\beta X_j} = \prod_j RX(2\beta)$, where $RX(\theta) = e^{-i\theta X/2}$ denotes a single-qubit rotation about the $X$-axis.

As we will refer to these later, we note that the Pauli-$Y$ operator is \[ Y = \begin{bmatrix} 0 & -i \\ i & 0 \end{bmatrix}, \] and the Hadamard gate \[ H = \tfrac{1}{\sqrt{2}}\begin{bmatrix}1&1\\1&-1\end{bmatrix} \] maps the computational basis state $\ket{0}$ to the equal superposition $\ket{+} = H \ket{0} = \tfrac{1}{\sqrt{2}}(\ket{0}+\ket{1})$, and is used throughout to initialize qubits in an unbiased state. 

The initial state for QAOA is chosen to be an eigenstate of the mixer $H_B$.
Specifically, we choose $\ket{s} = \ket{+}^{\otimes n} = \frac{1}{\sqrt{2^n}}\sum_{x \in \{0, 1\}^n} \ket{x}$, the uniform superposition over all computational basis states.
Then the QAOA ansatz at depth $p$ is \[
    \ket{\vec{\gamma}, \vec{\beta}} =
    \prod_{i=1}^{p} U_B(\beta_i)\, U_f(\gamma_i)\, \ket{s}.
\] The angles $\vec{\gamma} = (\gamma_1,\dots,\gamma_p)$ and $\vec{\beta} = (\beta_1,\dots,\beta_p)$ are real-valued parameters that are classically chosen (typically via gradient descent methods) to minimize the expected value of the cost Hamiltonian, \[
    \langle H_f \rangle =
    \bra{\vec{\gamma}, \vec{\beta}} H_f \ket{\vec{\gamma}, \vec{\beta}}.
\] The quality of the resulting solution is characterized by the approximation ratio
\begin{equation*}
  \mathrm{AR} = \frac{\langle H_f \rangle - C_{\max}}{C_{\min} - C_{\max}},
\end{equation*}
where $C_{\min}$ and $C_{\max}$ are the minimum and maximum eigenvalues of $H_f$.
By construction, $\mathrm{AR} \in [0,1]$ with $\mathrm{AR}=1$ at the global optimum.

Multi-angle QAOA (ma-QAOA) is a generalization of QAOA that introduces additional variational parameters and often achieves improved approximation ratios with fewer layers \cite{herrman2022multi, gaidai2024performance}.
Writing the cost and mixing Hamiltonians as sums of individual terms, 
\[
H_f = \sum_j H_f^{(j)}, \qquad H_B = \sum_k H_B^{(k)},
\] 
where $H_f^{(j)}$ ($H_B^{(k)}$) is the $j$-th ($k$-th) term of $H_f$ ($H_B$), ma-QAOA assigns an independent angle to each term, replacing the standard unitaries with \[ U_f(\vec{\gamma}) = e^{-i \sum_j \gamma_j H_f^{(j)}}, \qquad U_B(\vec{\beta}) = e^{-i \sum_k \beta_k H_B^{(k)}}.
\] This increased parameterization provides greater flexibility in shaping the variational landscape, at the cost of a higher-dimensional classical optimization problem.

\section{Existing Methods for Handling Constraints in Quantum Optimization}\label{sec: methods}
Section~\ref{sec: penalty-based qaoa} introduces penalty-based QAOA, which penalizes constraints in the cost Hamiltonian, while Section~\ref{sec: qaoa-plus} introduces feasibility-preserving mixing Hamiltonians (QAOA+ \cite{Hadfield_2019}). Finally, Section~\ref{sec: gm-qaoa} introduces Grover-mixer QAOA \cite{bartschi2020grover}, a specialized variant of QAOA+.

\subsection{Penalty-Based QAOA}\label{sec: penalty-based qaoa}
Penalty-based QAOA, which we term PenaltyQAOA, applies the reformulation of Sec.~\ref{sec:cop} directly as the QAOA cost Hamiltonian, yielding
\begin{equation*}
H_f^{\mathrm{pen}} = H_f + \sum_{k=1}^{|\mathcal{C}|} \delta_k\, H_{c_k},
\end{equation*}
where $H_{c_k}$ is the Hamiltonian encoding violation of constraint $c_k$~\cite{lucas2014ising, Hen_2016_a, Hen_2016_b}.
Despite its generality, this approach introduces well-known difficulties, such as large penalty coefficients distorting the energy landscape by compressing differences between feasible solutions while creating steep barriers at infeasible states, reducing the sensitivity of the cost Hamiltonian to objective improvements~\cite{mirkarimi2024quantum}.
Excessively large penalties can also induce highly oscillatory optimization landscapes~\cite{angara2025art, kulshrestha2022beinit, huembeli2021characterizing}, and no single penalty choice is guaranteed to be simultaneously effective across constraints with different natural scales.
PenaltyQAOA serves throughout this work as the primary baseline against which PC-QAOA is compared as it is commonly used to solve COPs in the literature.

\subsection{Quantum Alternating Operator Ansatz}\label{sec: qaoa-plus}
The Quantum Alternating Operator Ansatz (QAOA+) handles COPs by replacing the standard $X$-mixer with a problem-specific unitary $U_B(\beta)$ that preserves the feasible subspace~\cite{Hadfield_2019},
\[
\mathcal{H}_\mathcal{F} = \mathrm{span}\{\ket{x} : x \in \mathcal{F}\}, \qquad U_B(\beta)\,\mathcal{H}_\mathcal{F} \subseteq \mathcal{H}_\mathcal{F} \;\forall\,\beta.
\]
Starting from an initial feasible state $\ket{x} \in \mathcal{H}_\mathcal{F}$, the alternating ansatz of Sec.~\ref{sec:qaoa} then evolves entirely within $\mathcal{F}$, eliminating penalty terms and reducing the effective search space from $2^n$ to $|\mathcal{F}|$.
One may exploit problem structure directly through the mixer design, for example, $XY$-type mixers preserve fixed Hamming-weight constraints and swap-based mixers maintain permutation structure~\cite{Hadfield_2019}.

Designing an appropriate mixer for a general COP, however, requires explicit structural knowledge of $\mathcal{F}$~\cite{goldstein2024convergence}, and the initial feasible state $\ket{x}$ can itself be nontrivial to prepare for complex constraint sets~\cite{Hadfield_2019}.
The mixer must also induce sufficient connectivity within $\mathcal{F}$. If it cannot connect all feasible pairs, QAOA+ may be confined to a suboptimal region.
Furthermore, $XY$-mixer topologies with all-to-all connectivity or additional entangling gates can yield exponentially large dynamical Lie algebras, leading to barren plateaus~\cite{kordonowy2026lie}.

\subsection{Grover-Mixer QAOA}\label{sec: gm-qaoa}
Grover-Mixer QAOA (GM-QAOA) is a specialized variant of QAOA+ in which the feasible subspace is constructed explicitly via a state-preparation operator, and a global mixing operator is applied that preserves this subspace rather than acting on the full computational basis \cite{Hadfield_2019,bartschi2020grover,bridi2024analytical}.
The distinguishing feature of GM-QAOA is that the constraint dependence is placed primarily in the state-preparation operator rather than in a problem-specific local mixer \cite{bartschi2020grover}.

A GM-QAOA circuit of $p$ layers is comprised of the following components:
\begin{enumerate}
    \item State preparation using a unitary operator $U_S$ that creates an equal superposition of feasible states $\mathcal{F}$:
    \[ U_S\ket{0}^{\otimes n} = \ket{\mathcal{F}} := \frac{1}{\sqrt{|\mathcal{F}|}} \sum_{x \in \mathcal{F}} \ket{x}.
    \]
    \item Alternating applications of $p$ layers of the cost unitary $U_f(\gamma)=e^{-iH_f\gamma}$ and the Grover mixing unitary $U_B(\beta)=e^{-i\beta\ket{\mathcal{F}}\bra{\mathcal{F}}}$, resulting in the final state
    \[
        \ket{\vec{\gamma},\vec{\beta}}
        =
        \prod_{i=1}^p [U_B(\beta_i)U_f(\gamma_i)]\ket{\mathcal{F}}.
    \]
\end{enumerate}

GM-QAOA may therefore be viewed as a special case of feasibility-preserving QAOA in which the mixer is defined directly from the feasible-state superposition.
In contrast to approaches such as QAOA+, where one typically assumes that an initial feasible state is easy to prepare and then constructs a mixer that connects feasible solutions, GM-QAOA shifts this difficulty to preparing $\ket{\mathcal{F}}$ itself \cite{Hadfield_2019,bartschi2020grover}.
Indeed, the mixer can be written as \[ U_B(\beta) = e^{-i\beta\ket{\mathcal{F}}\bra{\mathcal{F}}} = U_S\left(I-(1-e^{-i\beta})\ket{0}\bra{0}\right)U_S^\dagger, \] which makes clear that implementing the mixing step reduces to implementing $U_S$ and $U_S^\dagger$ together with a selective phase on $\ket{0}^{\otimes n}$ \cite{bartschi2020grover}.
Thus, GM-QAOA can be applied whenever an efficient feasible-state preparation operator $U_S$ is available for the constraint family of interest~\cite{bartschi2020grover,bridi2024analytical}.
Providing a general-purpose construction of $U_S$ for arbitrary small-support constraints is one of the primary contributions of this work, which we address in Sec.~\ref{sec: vcg} via VCGs.
Notably, GM-QAOA initialized with an equal superposition of feasible states has been proven to avoid barren plateaus, with gradient variance scaling polynomially in system size \cite{tsvelikhovskiy2025provable}.

\section{Constraint Gadgets for Search Space Restriction}\label{sec: vcg}

In this section we introduce \emph{constraint gadgets} as a general mechanism for incorporating structural information into quantum optimization algorithms.
A constraint gadget prepares a quantum state supported only on solutions satisfying a selected constraint, thereby restricting the search space explored by the algorithm.

This perspective generalizes existing approaches such as GM-QAOA and QAOA+, which assume access to a feasible-state preparation operator $U_S$ but do not provide a general construction.
We consider both efficient constructions, which are available for certain classes of constraint families, and approximate constructions based on variational circuits that can be used for general constraints.

Let $(c_k, b_k)$ be a constraint with support $\mathrm{supp}(c_k)$ and feasible set $\mathcal{F}_k \subseteq \{0,1\}^{|\mathrm{supp}(c_k)|}$. 
A constraint gadget is a unitary $U_{c_k}$ such that
\begin{equation}
U_{c_k}\ket{0}^{\otimes |\mathrm{supp}(c_k)|}
=
\frac{1}{\sqrt{|\mathcal{F}_k|}} \sum_{x \in \mathcal{F}_k} \ket{x}.
\end{equation}
Constraint gadgets can be viewed as quantum analogues of classical domain reduction techniques, restricting the search space prior to optimization.

Ideally, we would like to apply a gadget for every constraint, however as each constraint gadget is implemented independently, some conflict can arise, meaning that applying two gadgets may not necessarily result in a uniform superposition of solutions satisfying both constraints. This is demonstrated in Appendix~\ref{app:overlapping_stateprep}.
The following proposition formalizes when gadgets compose without conflict.

\begin{proposition}[Gadget Composition]\label{prop:composition}
Let $(c_1, b_1),\ldots,(c_m,b_m)$ be constraints with pairwise disjoint supports $\mathrm{supp}(c_1),\ldots,\mathrm{supp}(c_m)$ that partition the variable indices, and let $|\mathcal{F}_k\rangle = \frac{1}{\sqrt{|\mathcal{F}_k|}}\sum_{x\in\mathcal{F}_k}|x\rangle$ denote the uniform superposition over the feasible set of $c_k$.
Then:
\begin{enumerate}
    \item[(i)] The joint feasible set factorizes: $\mathcal{F} = \mathcal{F}_1\times\cdots\times\mathcal{F}_m$.
    \item[(ii)] The uniform superposition over $\mathcal{F}$ decomposes as a tensor product: $|\mathcal{F}\rangle = |\mathcal{F}_1\rangle\otimes\cdots\otimes|\mathcal{F}_m\rangle$.
    \item[(iii)] Any circuit $V_k$ preparing $|\mathcal{F}_k\rangle$ on qubits $\mathrm{supp}(c_k)$ acts on a distinct register; the joint state $|\mathcal{F}\rangle$ is prepared by running all $m$ circuits in parallel.
\end{enumerate}
\end{proposition}
\begin{proof}
(i)~Since the supports are disjoint, satisfaction of $c_k$ depends only on variables in $\mathrm{supp}(c_k)$.
Thus $x\in\mathcal{F}$ iff $x|_{\mathrm{supp}(c_k)}\in\mathcal{F}_k$ for all $k$, which is the definition of $\mathcal{F}_1\times\cdots\times\mathcal{F}_m$. (ii)~The uniform superposition over a Cartesian product is the tensor product of the uniform superpositions over each factor. (iii)~The circuits $V_k$ act on disjoint qubit registers and therefore commute; they may be applied in any order or simultaneously.
\end{proof}

\subsection{Efficiently Generated Structured Constraint Gadgets}

While these circuits in theory prepare the feasible-state superposition exactly, gate error and noise prevent this in practice. We call these circuits exact to distinguish them from the variational gadgets of the next section.
Examples of constraint families admitting efficient exact constructions include:
\begin{itemize}

    \item \textbf{Cardinality constraints} ($\sum_i x_i = k$), which admit Dicke-state constructions~\cite{bartschi2019deterministic, bartschi2022short}.
    \item \textbf{Threshold constraints} ($\sum_i x_i \le k$ or $\ge k$), which can be expressed as superpositions of Dicke states~\cite{bartschi2019deterministic, bartschi2022short}.
    \item \textbf{Clique constraints} ($\sum_{v \in C} x_v \le 1$), which is a special case of threshold constraints and can be implemented in the same manner.
\end{itemize}

This work introduces a new method of implementing flow conservation constraints of the form $\sum_i x_i^{\mathrm{in}} = \sum_j x_j^{\mathrm{out}}$, which couple two variable sets and can be prepared using correlated Dicke-state constructions that enforce equal Hamming weight across the two registers.  The details of this construction can be found in Appendix~\ref{app:flow}.
These constructions exploit symmetry and typically require only logarithmic circuit depth in the support size, with no additional ancilla qubits~\cite{bartschi2019deterministic, bartschi2022short}.

\subsection{Variational Constraint Gadgets}

VCGs approximate feasible-state preparation for constraints lacking efficient exact constructions, using parameterized quantum circuits.

Let $(c_k, b_k)$ be a constraint with support $\mathrm{supp}(c_k)$ and feasible set $\mathcal{F}_k$.
We define a diagonal Hamiltonian $H_{C_k}$ on $|\mathrm{supp}(c_k)|$ qubits
\begin{equation}
H_{C_k}\ket{x} =
\begin{cases}
-\ket{x}, & x \in \mathcal{F}_k, \\
+\ket{x}, & x \notin \mathcal{F}_k,
\end{cases}
\end{equation}
so that each feasible state is a ground state, or lowest-energy eigenstate, of $H_{C_k}$ with eigenvalue $-1$, and each infeasible state is an excited state with eigenvalue $+1$.
Then minimizing $\langle H_{C_k} \rangle$ is equivalent to maximizing the total probability mass assigned to feasible states, and the optimal state is any superposition supported entirely on $\mathcal{F}_k$.
The Pauli-$Z$ expansion of $H_{C_k}$ is obtained via the Walsh--Hadamard transform~\cite{hadfield2021boolean, georges2025pauli}, requiring $\mathcal{O}(2^{|\mathrm{supp}(c_k)|}|\mathrm{supp}(c_k)|)$ classical time, which allows for easy implementation into a quantum circuit.
We use QAOA to optimize $\vec{\gamma}, \vec{\beta}$ by minimizing $\langle H_{C_k} \rangle$, producing a state
\[
\ket{\psi_k} = V(\vec{\gamma}, \vec{\beta})\ket{+}^{\otimes |\mathrm{supp}(c_k)|} \approx \ket{\mathcal{F}_k}.
\]
Note that QAOA may not be able to prepare the exact state at a given depth $p$, hence the approximation, with more layers yielding a better result.
Alternative approaches for preparing the ground state of $H_{C_k}$ include Grover's search algorithm~\cite{grover1997quantum} or QCOpt~\cite{nagarajan2021quantumcircuitopt}, which can achieve short circuit depth for small constraint supports at the cost of additional classical optimization.

We evaluate VCG performance using two complementary metrics.
The first is the \emph{feasible probability}
\begin{equation}\label{eq:feasible_prob}
P_{\mathcal{F}_k} = \frac{1 - \langle H_{C_k} \rangle}{2} = \Pr[x \in \mathcal{F}_k],
\end{equation}
which measures the total probability mass assigned to feasible states.
Note that this is equivalent to the approximation ratio of the VCG since each feasible state has eigenvalue $-1$ and each infeasible state has eigenvalue $+1$, implying that any $\mathrm{AR} < 1$ results in a final state with some amplitudes on infeasible states.
The second is the normalized feasible entropy, inspired by Shannon entropy~\cite{cover2006elements},
\begin{equation}\label{eq:feasible_entropy}
\mathcal{S}_{\mathrm{norm}} = \frac{-\sum_{x \in \mathcal{F}_k} q_x \log q_x}{\log |\mathcal{F}_k|},
\end{equation}
where $q_x = p(x)/P_{\mathcal{F}_k}$ is the measurement probability conditioned on feasibility, which measures how uniformly amplitude is distributed over feasible states (equal to 1 for a uniform distribution, 0 for a point mass).
Together, these metrics capture both feasibility and coverage of the feasible subspace.
For structured constraint families with exact constructions (e.g., Dicke-state-based gadgets), the resulting state achieves $P_{\mathcal{F}_k} = 1$ and $\mathcal{S}_{\mathrm{norm}} = 1$, corresponding to a uniform superposition over all feasible assignments.

Training is performed offline for each constraint type, producing reusable gadgets that can be incorporated into larger circuits without additional optimization.
Fig.~\ref{fig:gadget_flowchart} illustrates the full VCG construction pipeline for a representative knapsack constraint.
Further details on circuit design and training procedures are provided in Appendix~\ref{app:vcg}.

\begin{figure*}[t]
    \centering
    \resizebox{0.92\textwidth}{!}{
%
\begin{tikzpicture}[
    node distance=0.6cm and 0.5cm,
    >=Stealth,
    Hgate/.style={shading=axis, left color=rpLightBlue!65, right color=rpLightBlue!25, shading angle=135},
    Bgate/.style={shading=axis, left color=rpPink!70,      right color=rpPink!28,      shading angle=135},
    Mgate/.style={shading=axis, left color=rpBlue!65,      right color=rpBlue!25,      shading angle=135},
    probbox/.style={
        fill=rpBlue!20, draw=rpBlue!60!black,
        rounded corners, align=center, inner sep=9pt, font=\small
    },
    partbox/.style={
        fill=rpPink!22, draw=rpPink!55!black,
        rounded corners, align=center, inner sep=9pt, font=\small
    },
    strbox/.style={
        fill=rpGreen!22, draw=rpGreen!55!black,
        rounded corners, align=center, inner sep=9pt, font=\small
    },
    penbox/.style={
        fill=rpYellow!30, draw=rpYellow!60!black,
        rounded corners, align=center, inner sep=9pt, font=\small
    },
    circbox/.style={
        fill=rpLightBlue!28, draw=rpLightBlue!60!black,
        rounded corners, inner sep=4pt,
    },
    outbox/.style={
        fill=rpGreen!22, draw=rpGreen!55!black,
        rounded corners, align=center, inner sep=9pt, font=\small
    },
    targetbox/.style={
        fill=rpLightBlue!22, draw=rpLightBlue!55!black,
        rounded corners, align=center, inner sep=9pt, font=\small, dashed
    },
    arr/.style={->, thick, draw=rpBlack},
    lbl/.style={anchor=north west, inner sep=3pt, font=\scriptsize\bfseries},
]

\node[probbox, align=center, ] (constraint) {%
    \textbf{Constraint:}\quad
    $6x_3 + 2x_4 + 2x_5 \leq 3$\\[4pt]
    \begin{tabular}{@{}ll@{}}
      Feasible ($x$ satisfies constraint): & $\{000,\;001,\;010\}$\\[2pt]
      Infeasible ($x$ violates constraint): & $\{011,\;100,\;101,\;110,\;111\}$
    \end{tabular}%
};
\node[lbl] at (constraint.north west) {(a)};

\node[partbox, right=0.6cm of constraint, align=center, ] (labeled) {%
    \textbf{Feasibility Classification}\;$\ket{x_3\,x_4\,x_5}$\\[5pt]
    \begin{tabular}{@{}c@{\qquad}c@{}}
      \textbf{Feasible} ($x \in \mathcal{F}_k$)
        & \textbf{Infeasible} ($x \notin \mathcal{F}_k$)\\[2pt]
      \textcolor{blue}{\footnotesize$\ket{000},\;\ket{001},\;\ket{010}$}
        & \textcolor{red}{\footnotesize$\ket{011},\;\ket{100},$}\\[1pt]
        &\textcolor{red}{\footnotesize$\ket{101},\;\ket{110},\;\ket{111}$}
    \end{tabular}%
};
\node[lbl] at (labeled.north west) {(b)};

\node[penbox, below=0.6cm of labeled, align=center, ] (diagonal) {%
    \textbf{Set Diagonal of $H_{C_k}$:}\quad
    $d(x) = \begin{cases} -1 & x \in \mathcal{F}_k \\ +1 & x \notin \mathcal{F}_k \end{cases}$
    \\{\footnotesize(entries of the $8{\times}8$ diagonal Hamiltonian matrix)}%
};
\node[lbl] at (diagonal.north west) {(c)};

\node[strbox, left=0.6cm of diagonal, ] (hamiltonian) {%
    \textbf{Walsh--Hadamard Transform}
    $\;\rightarrow\;$
    Pauli-$Z$ decomposition of $H_{C_k}$:\\[5pt]
    \begin{tabular}{@{}r@{\;}l@{}}
      $H_{C_k}$
        & $= \tfrac{1}{4}I
            - \tfrac{3}{4}Z_{x_3}
            - \tfrac{1}{4}Z_{x_4}
            - \tfrac{1}{4}Z_{x_5}$\\[3pt]
        & $\phantom{{}={}}{}
            + \tfrac{1}{4}Z_{x_4}Z_{x_5}
            - \tfrac{1}{4}Z_{x_3}Z_{x_5}
            - \tfrac{1}{4}Z_{x_3}Z_{x_4}$\\[3pt]
        & $\phantom{{}={}}{}
            + \tfrac{1}{4}Z_{x_3}Z_{x_4}Z_{x_5}$
    \end{tabular}%
};
\node[lbl] at (hamiltonian.north west) {(d)};

\node[circbox, below=0.6cm of hamiltonian] (circuit) {%
    \begin{tabular}{c}
    \textbf{ma-QAOA Gadget Circuit}\quad$V(\vec\gamma,\vec\beta)$\\[6pt]
    {
    \begin{quantikz}[row sep=0.7em, column sep=0.55em]
        \lstick{$x_3$} & \gate[style={Hgate}]{H} & \gate[wires=3, style={Bgate}]{U_{C_k}(\vec\gamma)} & \gate[style={Mgate}]{RX(2\beta_3)} & \qw \\
        \lstick{$x_4$} & \gate[style={Hgate}]{H} & & \gate[style={Mgate}]{RX(2\beta_4)} & \qw \\
        \lstick{$x_5$} & \gate[style={Hgate}]{H} & & \gate[style={Mgate}]{RX(2\beta_5)} & \qw
    \end{quantikz}%
    }\\[4pt]
    {\footnotesize $U_{C_k}(\vec\gamma)$: one phase gate per Pauli term (7 non-identity terms)}
    \end{tabular}%
};
\node[lbl] at (circuit.north west) {(e)};

\node[penbox, right=0.9cm of circuit, ] (training) {%
    \textbf{Training Procedure}\\[5pt]
    \begin{tabular}{@{}l@{}}
      \textbf{1.} QAOA $p{=}1$ (shared $\gamma,\beta$)
                  $\;\Rightarrow\;$ warm-start angles $(\gamma^*,\beta^*)$\\[4pt]
      \textbf{2.} ma-QAOA $p{=}1$: seed angles from $(\gamma^*,\beta^*)$\\[4pt]
      \textbf{3.} ma-QAOA $p{>}1$: warm-start from depth $p{-}1$ optimum\\[4pt]
      \textbf{Stop} when
          $P_{\mathcal{F}_k} = \tfrac{1-\langle H_{C_k}\rangle}{2} \;\geq\; \tau$\\[3pt]
      \hphantom{\textbf{Stop} }AND\;
          $\mathcal{S}_{\mathrm{norm}} = \tfrac{-\sum_{x\in\mathcal{F}_k} q_x\log q_x}{\log|\mathcal{F}_k|} \;\geq\; \eta$
    \end{tabular}%
};
\node[lbl] at (training.north west) {(f)};


\coordinate (branch) at ($(training.east)+(3.95cm,0)$);

\node[outbox] (gadgetdb)
    at ($(branch)+(0,1.8cm)$) {%
    \textbf{Gadget Stored in DB}\\[4pt]
    \begin{tabular}{@{}l@{}}
      Optimal $\vec\gamma^*,\,\vec\beta^*$\\[2pt]
      Depth $p^*$, AR, $\mathcal{S}_{\mathrm{norm}}$\\[2pt]
      Reusable for future COPs
    \end{tabular}%
};
\node[lbl] at (gadgetdb.north west) {(g)};

\node[targetbox] (target)
    at ($(branch)+(0,-1.75cm)$) {%
    \textbf{Approximate Ground State}\;$\ket{\psi_k}\approx\ket{\mathcal{F}_k}$\\[5pt]
    $\dfrac{1}{\sqrt{3}}\bigl(\ket{000}+\ket{001}+\ket{010}\bigr)$\\[4pt]
    {\footnotesize Equal superposition of all 3 feasible states; no ancilla}%
};
\node[lbl] at (target.north west) {(h)};

\draw[arr] (constraint)  -- (labeled);
\draw[arr] (labeled)     -- (diagonal);
\draw[arr] (diagonal)    -- (hamiltonian);
\draw[arr] (hamiltonian) -- (circuit);
\draw[arr] (circuit)     -- (training);

\draw[arr] (training.east) -- (branch);
\draw[arr] (branch) -- (gadgetdb.south);
\draw[arr] (branch) -- (target.north);

\coordinate (loopleft)       at ($(circuit.south)+(0,-1.3cm)$);
\coordinate (loopleftbottom) at (loopleft -| training.south);
\draw[arr] (training.south) -- (loopleftbottom)
    -- node[above, font=\footnotesize, align=right, xshift=1.2cm]
        {$P_{\mathcal{F}_k}<\tau$ and $\mathcal{S}_{\mathrm{norm}}<\eta$:\\[-2pt] $p \rightarrow p+1$,\\[-2pt] warm-start}
    (loopleft) -- (circuit.south);

\end{tikzpicture}}
    \caption{VCG construction and training pipeline for the knapsack constraint
    $6x_3 + 2x_4 + 2x_5 \leq 3$.
    \textbf{(a)}~Constraint definition and feasible/infeasible classification.
    \textbf{(b)}~Labeling of computational basis states.
    \textbf{(c)}~Construction of the diagonal constraint Hamiltonian $H_{C_k}$.
    \textbf{(d)}~Walsh--Hadamard transform yields the Pauli-$Z$ decomposition of $H_{C_k}$.
    \textbf{(e)}~Parameterized ma-QAOA circuit $V(\vec{\gamma},\vec{\beta})$ trained to minimize $\langle H_{C_k}\rangle$.
    \textbf{(f)}~Layer-wise training loop: warm-started ma-QAOA with early stopping when $P_{\mathcal{F}_k} \geq \tau$ and $\mathcal{S}_{\mathrm{norm}} \geq \eta$.
    \textbf{(g)}~Converged gadget stored in the reusable gadget library.
    \textbf{(h)}~Approximate ground state $|\psi_k\rangle \approx |\mathcal{F}_k\rangle$, ready for use as a structural state-preparation component.}
    \label{fig:gadget_flowchart}
\end{figure*}

\subsection{Integration Into Quantum Optimization Algorithms}

For a collection of independent constraints, the joint state is
\begin{equation}
\ket{\psi} = \bigotimes_{k} \ket{\psi_k} \approx \bigotimes_{k} \ket{\mathcal{F}_k},
\end{equation}
which lies approximately in the intersection of the individual feasible subspaces.
This state can be used directly in GM-QAOA or other feasibility-preserving frameworks, such as QAOA+ with $XY$ mixers, confining most amplitude to the desired subspace throughout the evolution.

When gadgets are approximate, the joint feasible probability of the composed state degrades with the number of gadgets, but the following corollary guarantees that it remains high as long as each gadget individually achieves high feasible probability.

\begin{corollary}[Error Accumulation]\label{cor:error}
Under the same conditions as Proposition~\ref{prop:composition}, if each gadget prepares a state with feasible probability $P_{\mathcal{F}_k} \geq 1-\epsilon_k$, then the composed state satisfies
\[
P_{\mathcal{F}}
\;\geq\; \prod_{k=1}^{|\mathcal{C}|}(1-\epsilon_k) \;\geq\; 1-\sum_{k=1}^{|\mathcal{C}|}\epsilon_k.
\]
\end{corollary}
Since all VCGs in our experiments achieve $P_{\mathcal{F}_k}\geq 0.9999$, the per-gadget infeasible leakage satisfies $\epsilon_k\leq 10^{-4}$, so Corollary~\ref{cor:error} guarantees that the joint feasible probability across all structural constraints exceeds $1 - |\mathcal{C}|\cdot 10^{-4}$.
When constraints have overlapping supports, the joint feasible set does not factorize and sequential application of individual gadgets introduces interference. This case is treated in Appendix~\ref{app:overlapping_stateprep}.

\subsection{Computational Considerations and Limitations}

The primary limitation of VCGs is the exponential dependence on constraint support size.
Constructing $H_{C_k}$ requires enumerating $2^{|\mathrm{supp}(c_k)|}$ assignments, and its Pauli decomposition may contain up to $2^{|\mathrm{supp}(c_k)|}$ terms.

In practice, this limits VCGs to small-support constraints (e.g., $|\mathrm{supp}(c_k)| \le 5$), where circuit depth and training remain manageable.
For larger constraints, classical preprocessing techniques such as variable fixing via bound propagation or symmetry-based aggregation \cite{Margot2010} may reduce effective support prior to gadget construction. 

 While high feasible probability can be achieved, imperfect training may result in non-uniform amplitude distributions over $\mathcal{F}_k$, concentrating amplitude on a subset of feasible states rather than the full equal superposition.
 Consequently, VCGs are most appropriate when exact constructions are unavailable, reinforcing the hybrid approach developed in the next section.

An additional consideration is the storage and reuse of trained constraint gadgets.
In practice, VCGs can be stored in a library indexed by constraint structure and parameters, enabling reuse across problem instances and spreading the cost of offline training across many instances.
This is particularly effective for recurring constraint types (e.g., small knapsack or cardinality constraints), where a single trained gadget can be applied repeatedly.

\section{PC-QAOA: Partitioned-Constraint QAOA}\label{sec: hybrid}

The approaches discussed in Sections~\ref{sec: penalty-based qaoa}--\ref{sec: gm-qaoa} represent two extremes in constraint handling for quantum optimization: fully energetic enforcement via penalties, and fully structural enforcement via feasible-subspace state preparation and custom mixers.
We introduce \emph{PC-QAOA}, a framework that interpolates between these paradigms by partitioning constraints according to whether they are enforced structurally or energetically.

Consider a constrained optimization problem \[
    \min_{x \in \{0,1\}^n} f(x)
    \quad
    \text{subject to}
    \quad
    c_k(x) \le b_k, \quad (c_k,b_k) \in \mathcal C. \] We partition the constraint set as $\mathcal C = \mathcal C_{\mathrm{str}} \cup \mathcal C_{\mathrm{pen}}$, where $\mathcal C_{\mathrm{str}}$ denotes constraints enforced structurally and $\mathcal C_{\mathrm{pen}}$ denotes those enforced via penalty terms.
    The overall pipeline is illustrated in Fig.~\ref{fig:constraintCircuit}.

\begin{figure*}[t]
    \centering
    \resizebox{0.8\textwidth}{!}{
%
\begin{tikzpicture}[
    node distance=0.6cm and 0.3cm,
    >=Stealth,
    probbox/.style={
        fill=rpBlue!20, draw=rpBlue!60!black,
        rounded corners, align=center, inner sep=8pt, font=\small
    },
    partbox/.style={
        fill=rpPink!22, draw=rpPink!55!black,
        rounded corners, align=center, inner sep=7pt, font=\small
    },
    abox/.style={   
        fill=rpGreen!28, draw=rpGreen!55!black,
        rounded corners, align=center, inner sep=7pt, font=\small
    },
    bbox/.style={   
        fill=rpPink!30, draw=rpPink!60!black,
        rounded corners, align=center, inner sep=7pt, font=\small
    },
    cbox/.style={   
        fill=rpYellow!38, draw=rpYellow!65!black,
        rounded corners, align=center, inner sep=7pt, font=\small
    },
    circbox/.style={
        fill=rpLightBlue!28, draw=rpLightBlue!60!black,
        rounded corners, inner sep=3pt,
    },
    outbox/.style={
        fill=rpPink!20, draw=rpPink!50!black,
        rounded corners, align=center, inner sep=7pt, font=\small
    },
    retbox/.style={
        fill=rpGreen!25, draw=rpGreen!60!black,
        rounded corners, align=center, inner sep=7pt, font=\small
    },
    arr/.style={->, thick, draw=rpBlack},
    lbl/.style={anchor=north west, inner sep=3pt, font=\scriptsize\bfseries},
    Agate/.style={shading=axis, left color=rpGreen!65, right color=rpGreen!25, shading angle=135},
    Bgate/.style={shading=axis, left color=rpPink!70,  right color=rpPink!28,  shading angle=135},
    Cgate/.style={shading=axis, left color=rpYellow!75,right color=rpYellow!32,shading angle=135},
    Fgate/.style={shading=axis, left color=rpRed!55,right color=rpRed!18,shading angle=135},
]

\node[probbox] (problem) {%
    \textbf{Constrained Optimization Problem}\\[5pt]
    $\displaystyle\min_{x\in\{0,1\}^7}\; x^\top Q\, x$
    \qquad
    \begin{tabular}{@{}r@{\;}l@{\quad}l@{}}
      \textbf{A:}              & $x_0+x_1+x_2 = 1$         & support $\{0,1,2\}$\\[2pt]
      \textbf{B:}              & $6x_3+2x_4+2x_5 \le 3$    & support $\{3,4,5\}$\\[2pt]
      \textbf{C:}              & $x_1+x_4+x_6 \le 1$       & support $\{1,4,6\}$ (overlapping)
    \end{tabular}
};
\node[lbl] at (problem.north west) {(a)};

\node[partbox, below=of problem, text width=13.5cm] (partition) {%
    \textbf{Classify Constraints into Structural and Penalty}\\[4pt]
    \begin{tabular}{@{}ll@{}}
      \emph{Structural:} & constraints admitting exact circuit prep (Dicke) or a trainable VCG gadget\\[2pt]
      \emph{Penalty:}    & constraints with overlapping support — enforced energetically via $U^C_{\text{pen}}$
    \end{tabular}
};
\node[lbl] at (partition.north west) {(b)};

\node[bbox, below=0.9cm of partition, text width=4.0cm] (vcg) {%
    \textbf{B\;—\;VCG}\\[4pt]
    \begin{tabular}{@{}l@{}}
      Weighted inequality\\[2pt]
      $\Rightarrow$ train ma-QAOA on $H_{C_k}$\\
      \quad over $\{x_3,x_4,x_5\}$\\[2pt]
      $\Rightarrow$ Grover mixer\\[4pt]
      
    \end{tabular}%
};
\node[lbl] at (vcg.north west) {(d)};

\node[abox, left=of vcg, text width=4.0cm] (dicke) {%
    \textbf{A\;—\;Exact Dicke Prep}\\[4pt]
    \begin{tabular}{@{}l@{}}
      Unit coefficients, equality\\[2pt]
      $\Rightarrow$ log-depth W-state circuit\\[2pt]
      $\Rightarrow$ XY mixer on $\{x_0,x_1,x_2\}$\\[4pt]
    \end{tabular}%
};
\node[lbl] at (dicke.north west) {(c)};

\node[cbox, right=of vcg, text width=4.0cm] (pen) {%
    \textbf{C\;—\;Penalty Term}\\[4pt]
    \begin{tabular}{@{}l@{}}
      Overlaps both A and B\\[2pt]
      $\Rightarrow$ add slack $s_C$\\[2pt]
      $\Rightarrow$ $U^C_{\text{pen}}(\lambda\gamma)$\\
      \quad on $\{x_1,x_4,x_6,s_C\}$\\[4pt]
      \emph{Adds slack qubit $s_C$}
    \end{tabular}%
};
\node[lbl] at (pen.north west) {(e)};

\node[circbox, below=1.3cm of vcg] (circuit) {%
    \begin{tabular}{c}
    \textbf{PC-QAOA Circuit}\\[6pt]
    {
    \begin{quantikz}[row sep=0.75em, column sep=0.6em]
        \lstick{$x_0$}
            & \gate[wires=3, style={Agate}]{\text{Dicke}}
            & \gate[wires=6, style={Fgate}]{U_f(\gamma)}
            \gategroup[wires=7, steps=1, style={dashed, inner xsep=-3pt, inner ysep=-3pt}]{{}}
            \gategroup[wires=8, steps=4, style={dashed, thick, inner xsep=-1pt, inner ysep=-1pt, color=rpPink}]{$p$-times}
            & \qw
            & \qw
            & \gate[wires=3, style={Agate}]{\text{XY-Mixer}}
            & \meter{}
            \\
        \lstick{$x_1$}
            &
            &
            & \qw
            & \gate[style={Cgate}]{U^C_{\text{pen}}(\lambda\gamma)}\gategroup[wires=7, steps=1, style={dashed, inner xsep=-3pt, inner ysep=-3pt}]{}
            &
            & \meter{}
            \\
        \lstick{$x_2$}
            &
            &
            & \qw
            & \qw
            &
            & \meter{}
            \\
        \lstick{$x_3$}
            & \gate[wires=3, style={Bgate}]{\text{VCG}}
            &
            & \qw
            & \qw
            & \gate[wires=3, style={Bgate}]{\text{Grover}}
            & \meter{}
            \\
        \lstick{$x_4$}
            &
            &
            & \qw
            & \gate[style={Cgate}]{U^C_{\text{pen}}(\lambda\gamma)}
            &
            & \meter{}
            \\
        \lstick{$x_5$}
            &
            &
            & \qw
            & \qw
            &
            & \meter{}
            \\
        \lstick{$x_6$}
            & \gate[style={Fgate}]{H}
            & \gate[wires=1, style={Fgate}]{U_f(\gamma)}
            & \qw
            & \gate[style={Cgate}]{U^C_{\text{pen}}(\lambda\gamma)}
            & \gate[style={Cgate}]{RX(\beta)}
            & \meter{}
            \\
        \lstick{$s_C$}
            & \gate[style={Fgate}]{H}
            & \qw
            & \qw
            & \gate[style={Cgate}]{U^C_{\text{pen}}(\lambda\gamma)}
            & \gate[style={Cgate}]{RX(\beta)}
            & \meter{}
    \end{quantikz}
  }
    \\[5pt]
    \footnotesize
    \end{tabular}%
};
\node[lbl] at (circuit.north west) {(f)};

\node[outbox, right=0.9cm of circuit] (output) {%
    \textbf{Adaptive Training Loop}\\[6pt]
    \begin{tabular}{@{}r@{\;\;}l@{}}
      \textbf{1.} & Optimize angles at depth $p$ via Adam (random restarts) \\[3pt]
      \textbf{2.} & Sample bitstrings; evaluate $P(\text{feas})$
    \end{tabular}%
};
\node[lbl] at (output.north west) {(g)};

\node[retbox, below=1.4cm of output] (ret) {%
    \textbf{Return} best feasible $x^{*} \in \{0,1\}^7$
};
\node[lbl, xshift=-1pt, yshift=1pt] at (ret.north west) {(h)};

\coordinate (branch) at ($(partition.south)+(0,-0.35cm)$);
\draw[arr] (partition.south) -- (branch);
\draw[arr] (branch) -| (dicke.north);
\draw[arr] (branch) -- (vcg.north);
\draw[arr] (branch) -| (pen.north);

\coordinate (merge) at ($(circuit.north)+(0,0.4cm)$);
\draw[arr] (dicke.south) |- (merge);
\draw[arr] (vcg.south)   -- (merge);
\draw[arr] (pen.south)   |- (merge);
\draw[arr] (merge) -- (circuit.north);

\draw[arr] (problem.south) -- (partition.north);
\draw[arr] (circuit.east) -- (output.west);   
\draw[arr] (output.south)
    -- node[right, font=\footnotesize, align=left]
        {$p = p_{\max}$}
    (ret.north);     

\coordinate (looptop) at ($(output.north)+(0, 0.3cm)$);
\draw[arr]
    (output.north)
    -- (looptop)
    -- node[above, font=\footnotesize, align=center]
        {$p < p_{\max}$: $p \mathrel{+}= 1$, warm-start angles}
    (looptop -| circuit.east);

\end{tikzpicture}}
    \caption{PC-QAOA pipeline illustrating constraint partitioning, structural gadget preparation, and energetic enforcement.}
    \label{fig:constraintCircuit}
\end{figure*}

\subsection{Constraint Partitioning}

We use a greedy two-pass partitioning rule. In the first pass, constraints admitting exact analytic state preparation are greedily added to $\mathcal{C}_\mathrm{str}$ in order, provided their support is disjoint from all previously selected constraints. In the second pass, remaining constraints admitting a VCG are added to $\mathcal{C}_\mathrm{str}$ under the same disjointness condition. All remaining constraints are assigned to $\mathcal{C}_\mathrm{pen}$.
This ensures that all constraints in $\mathcal C_{\mathrm{str}}$ act on non-overlapping qubit registers.
By Proposition~\ref{prop:composition}, their joint feasible superposition is the tensor product of the individual gadget states and can be prepared in parallel, with joint infidelity bounded by $\sum_k \epsilon_k$ (Corollary~\ref{cor:error}).


\subsection{Structural Enforcement}

Denote the partially feasible region by
\[
\mathcal F_\mathrm{str}
=
\left\{
x\in\{0,1\}^n :
c_{k_s}(x)\leq b_{k_s}
\text{ for all } k_s = 1,\ldots,|\mathcal C_\mathrm{str}|
\right\}.
\]
Exact gadgets prepare the uniform superposition over $\mathcal{F}_{\mathrm{str}}$, while VCGs approximate this distribution.
In both cases, the ansatz is biased toward feasibility by restricting the accessible subspace.

\subsection{Energetic Enforcement of Residual Constraints}

The penalized cost Hamiltonian is
\begin{equation}
H_f^{\mathrm{PC}}
=
H_f
+
\sum_{k_p=1}^{|\mathcal C_{\mathrm{pen}}|}
\delta_{k_p} H_{c_{k_p}}.
\label{eq: hhybrid}
\end{equation}
Structural gadgets restrict the initial state to $\mathcal{F}_{\mathrm{str}}$. For exact gadgets, the mixer exactly preserves this subspace while  it approximately preserves for VCG-based gadgets. Consequently, both slack variables and penalty terms  only need to consider this reduced space.

Regarding slack variables, recall that the number of slacks required for constraint $(c_k,b_k)$ is a function of the difference between $b_k$ and $c_k^{\min}$. In this case, $c_k^{\min}$ can be computed over  $\mathcal{F}_{\mathrm{str}}$ instead of all binary vectors. This restriction can  increase  $c_k^{\min}$, potentially resulting in the need for fewer slack variables.

Regarding penalty parameters,  a sufficient condition for an appropriate penalty  is
\begin{equation}
\delta_k > \max_{x \in \mathcal{F}_{\mathrm{str}}} f(x) - \min_{x \in \mathcal{F}_{\mathrm{str}}} f(x).
\end{equation}

Since $\mathcal{F}_{\mathrm{str}} \subseteq \{0,1\}^n$, this range is typically much smaller than the full objective range, so PC-QAOA requires smaller penalty coefficients.
Large penalty coefficients cause the penalty terms to dominate the cost Hamiltonian, so the optimizer spends circuit depth suppressing infeasible states rather than distinguishing between feasible solutions by objective value.
PC-QAOA optimizes the $p$-layer ansatz
\begin{equation*}
|\psi_p(\vec{\gamma},\vec{\beta})\rangle
= \prod_{l=1}^{p}
\Bigl[U_M(\vec{\beta}_l)\,e^{-i\gamma_l H_f^{\mathrm{PC}}}\Bigr]
|\mathcal{F}_{\mathrm{str}}\rangle,
\end{equation*}
where $|\mathcal{F}_{\mathrm{str}}\rangle$ is prepared by the structural gadgets (Section~\ref{sec: vcg}), $H_f^{\mathrm{PC}}$ is defined in Eqn.~\eqref{eq: hhybrid}, and $U_M(\vec{\beta}_l)$ is the per-gadget mixer
\begin{equation*}
U_M(\vec{\beta}_l)
=
\bigotimes_{k_s=1}^{|\mathcal C_{\mathrm{str}}|} U_M^{(k_s)}(\beta_{k_s}^{(l)})
\;\otimes\;
\bigotimes_{j\,\notin\,\mathrm{supp}(\mathcal C_{\mathrm{str}})} e^{-i\beta_j^{(l)} X_j},
\end{equation*}
where $\mathrm{supp}(\mathcal{C}_{\mathrm{str}}) = \bigcup_{k_s=1}^{|\mathcal{C}_{\mathrm{str}}|} \mathrm{supp}(c_{k_s})$ is the union of supports of all structural constraints.
The gadget-specific mixer $U_M^{(k_s)}$ preserves the feasible subspace of constraint $k_s$.
Cardinality equality constraints use $XY$ mixers, which preserve Hamming weight.
Inequality, flow conservation, and variational gadgets use a per-gadget Grover mixer.
Qubits not covered by any structural gadget use standard $X$ mixers.
As with all QAOA variants, the variational parameters $(\vec{\gamma},\vec{\beta})$ are chosen to minimize the expected cost
\begin{equation*}
\mathcal{L}(\vec{\gamma},\vec{\beta})
= \langle\psi_p(\vec{\gamma},\vec{\beta})\,|\,H_f^{\mathrm{PC}}\,|\,\psi_p(\vec{\gamma},\vec{\beta})\rangle.
\end{equation*}
Because the dynamics are confined to $\mathcal{F}_{\mathrm{str}}$, this objective simultaneously minimizes the QUBO cost and penalizes violations of the residual constraints within the already-reduced feasible subspace.

\subsection{Trade-offs in Constraint Handling}
PC-QAOA interpolates between PenaltyQAOA where $\mathcal C_{\mathrm{str}}=\emptyset$ and fully structural approaches such as QAOA+ where $\mathcal C_{\mathrm{pen}}=\emptyset$. Moving constraints from $\mathcal C_{\mathrm{pen}}$ to $\mathcal C_{\mathrm{str}}$ reduces reliance on penalty terms at the cost of increased circuit complexity.
The partition $(\mathcal C_{\mathrm{str}}, \mathcal C_{\mathrm{pen}})$ is therefore chosen to balance structural enforcement with energetic flexibility. In addition to circuit complexity, the choice of $(\mathcal C_{\mathrm{str}}, \mathcal C_{\mathrm{pen}})$ will likely have a considerable impact on the number of iterations required for convergence. Motivated by the analysis of Grover's search algorithm, our intuition suggests that an effective constraint partition should eliminate as many infeasible solutions as possible by assigning them zero amplitude. Recall that Grover's search requires $O(\sqrt{F})$ iterations, where $F$ is the number of solutions with non-zero amplitude~\cite{grover1997quantum}. By analogy, reducing the size of the search space through structural enforcement  might reduce the number of iterations required for convergence.

\section{Results}\label{sec: Results}

PC-QAOA is evaluated across seven constraint families and problem sizes $n \in \{3,\ldots,8\}$.
The specific details of the problems tested can be found in Section~\ref{sec: instances}.
We first characterize the performance of VCGs in Section~\ref{sec: vcg-results}, demonstrating their ability to achieve high feasible probability and uniform coverage of the feasible subspace.
Next, we evaluate the full PC-QAOA pipeline in Section~\ref{sec: pcqaoa-reselts}, showing that it outperforms PenaltyQAOA on all metrics considered while also using fewer resources.

\subsection{Problem Instances}\label{sec: instances}

All simulations use PennyLane~\cite{bergholm2018pennylane} with JAX and \texttt{qjit} just-in-time compilation, and are run on classical hardware via exact statevector simulation.
The seven constraint families tested are summarized in Table~\ref{tab:constraint_families}. Exact structural circuit constructions for families not handled by VCGs are described in Appendix~\ref{app: structural encoding}.

\begin{table*}[t]
\centering
\caption{Constraint families used in experiments, with the number of times each appears as a structural (state-prep) or penalized constraint in PC-QAOA across all 500 problem instances. Counts reflect the greedy disjointness-first partition.}
\label{tab:constraint_families}
\begin{tabular}{llcccc}
\hline
\textbf{Family} & \textbf{Form} & \boldmath$|\mathrm{supp}(c_k)|$ & \textbf{Structural} & \textbf{Penalized} & \textbf{Handler} \\
\hline
Cardinality ($=$)           & $\sum_i x_i = k$                                         & 2--5 & 130 &  14 & Dicke state (exact) \\
Cardinality ($\leq$/$\geq$) & $\sum_i x_i \leq k$ / $\geq k$                          & 2--5 & 246 &  91 & Superposition of Dicke states (exact) \\
Assignment                  & $\sum_j x_{ij} = 1$                                      & 2--5 &  27 &   1 & Dicke state (exact) \\
Flow conservation           & $\sum_i x_i^{\mathrm{in}} = \sum_j x_j^{\mathrm{out}}$  & 2--5 &  56 &  20 & Flow state prep (exact) \\
Independent set            & $x_i x_j = 0$                                            & 2    &  14 &   2 & LEQ state prep (exact) \\
Knapsack                    & $\sum_i w_i x_i \leq k$                                  & 3--5 & 148 & 109 & VCG (pre-trained) \\
Quadratic knapsack          & $\sum_{ij} Q_{ij} x_i x_j \leq k$                       & 3--5 & 154 & 128 & VCG (pre-trained) \\
\hline
\textit{Total}              &                                                           &      & 775 & 365 & \\
\hline
\end{tabular}
\end{table*}

Random-coefficient quadratic objective functions are generated independently for each problem size $n \in \{3,\ldots,8\}$, with 10 possible objectives per $n$.
We create 500 COPs total, with each instance being an (objective, constraint set) pair with 2--3 constraints drawn from combinations of the seven families in Table~\ref{tab:constraint_families}.
If a constraint is assigned to the structural subset but lacks an exact construction and no pre-trained VCG is available in the gadget library, a VCG is trained on the fly before optimization proceeds. 
Table~\ref{tab:experiment_tasks} summarizes the two problem sets: a \emph{disjoint} set where all constraint variable supports are non-overlapping (all constraints handled structurally) and an \emph{overlapping} set where constraints share variables, forcing a structural/penalty split.

The COPs in this study are intentionally small ($n \leq 8$), reflecting the qubit budgets accessible on near-term quantum hardware.
The effective qubit count for PenaltyQAOA is larger still, since each penalized inequality constraint requires additional logarithmic slack qubits.
At this scale, classical solvers such as Gurobi \cite{gurobi} find exact optima in milliseconds, and exhaustive enumeration is feasible.
These experiments do not aim to demonstrate quantum advantage over classical methods, which requires problem sizes well beyond current hardware.
Instead, they validate two aspects of PC-QAOA: that structural constraint enforcement measurably improves feasibility and solution quality relative  to penalty-based QAOA, and that VCG-based preparation is as reliable as exact construction in practice.

\begin{table*}[t]
\centering
\caption{Experiment configurations for the PC-QAOA vs. PenaltyQAOA comparison. Each instance is a (QUBO, constraint set) pair with 2--3 constraints drawn from the seven families in Table~\ref{tab:constraint_families}. Disjoint instances have non-overlapping constraint supports (all constraints structural); overlapping instances have shared variables so at least one constraint is penalized. The smallest constraint family tested acts on 2 variables, so disjoint instances require at least two non-overlapping 2-variable blocks, giving $n \geq 4$; overlapping instances can share variables across constraints and therefore reach $n = 3$. Completed counts reflect runs that returned results; non-completions were due to transient XLA/JAX hardware errors on the compute cluster.}
\label{tab:experiment_tasks}
\begin{tabular}{lcccccc}
\hline
 & & & \multicolumn{2}{c}{\textbf{PC-QAOA}} & \multicolumn{2}{c}{\textbf{PenaltyQAOA}} \\
\textbf{Problem set} & \boldmath$n$ & \textbf{\# constraints} & \textbf{Attempted} & \textbf{Completed} & \textbf{Attempted} & \textbf{Completed} \\
\hline
Disjoint    & $4$--$8$ & 2--3 & 250 & 250 & 250 & 185 \\
Overlapping & $3$--$8$ & 2--3 & 250 & 163 & 250 & 131 \\
\hline
\textit{Total} & & & 500 & 413 & 500 & 316 \\
\hline
\end{tabular}
\end{table*}

\subsection{Constraint State Preparation}\label{sec: vcg-results}

We create VCGs for 30 knapsack and 30 quadratic-knapsack constraints.
Of these, 14 were found in the preprocessing step to have a feasible superposition with the same structure as a superposition of Dicke states, and the more efficient construction was used.
As no variational training is needed for these constraints, we assign $p=0$ to indicate this and they are not shown in the plots.
All other VCGs were trained using the procedure outlined in Fig.~\ref{fig:gadget_flowchart} with the probability of feasible threshold $\tau = 0.999$ and the normalized feasible entropy threshold $\eta = 0.9999$.
More information on how these VCGs are trained can be found in Table~\ref{tab:vcg_hyperparams}.

\begin{figure}[t]
  \centering
  \includegraphics[width=0.95\linewidth]{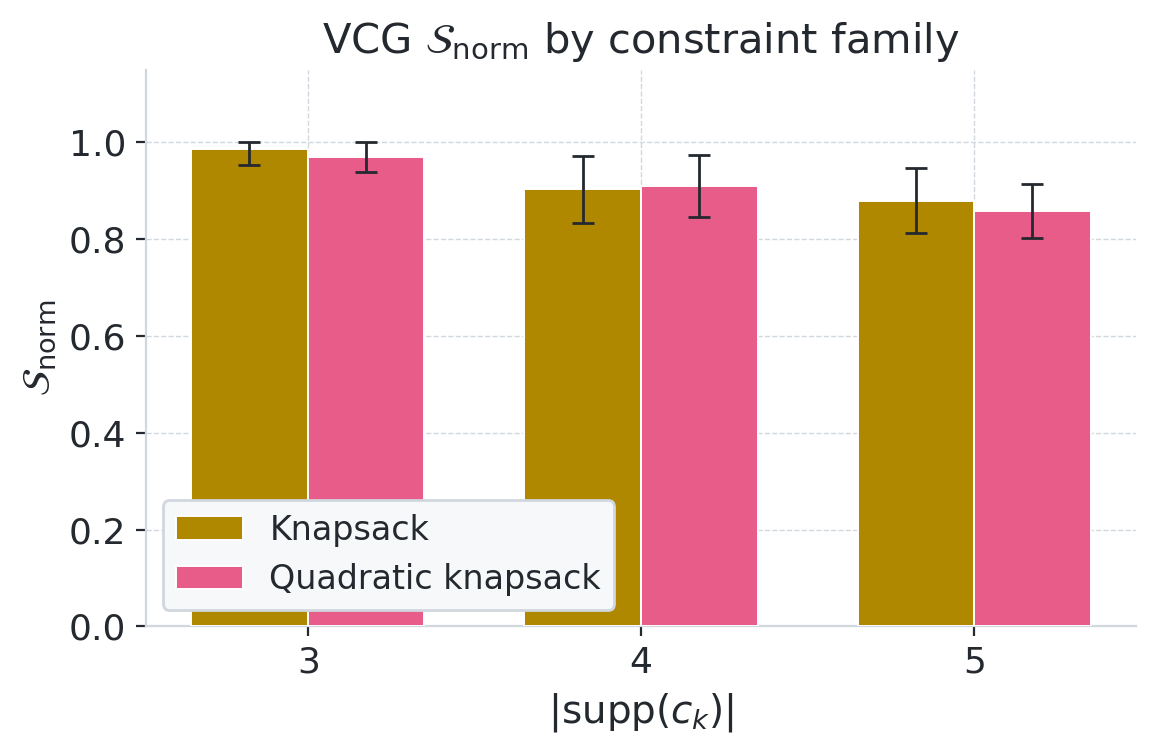}
  \caption{Normalized feasible entropy $\mathcal{S}_{\mathrm{norm}}$ of the trained VCG state, grouped by support size $|\mathrm{supp}(c_k)|$ and colored by constraint family. Error bars show one standard deviation. Knapsack gadgets achieve a higher mean normalized feasible entropy ($0.9204$) than quadratic knapsack ($0.9155$), indicating a more uniform spread of amplitude over feasible states.}
  \label{fig:vcg_entropy_by_type}
\end{figure}

\begin{figure}[t]
  \centering
  \includegraphics[width=0.95\linewidth]{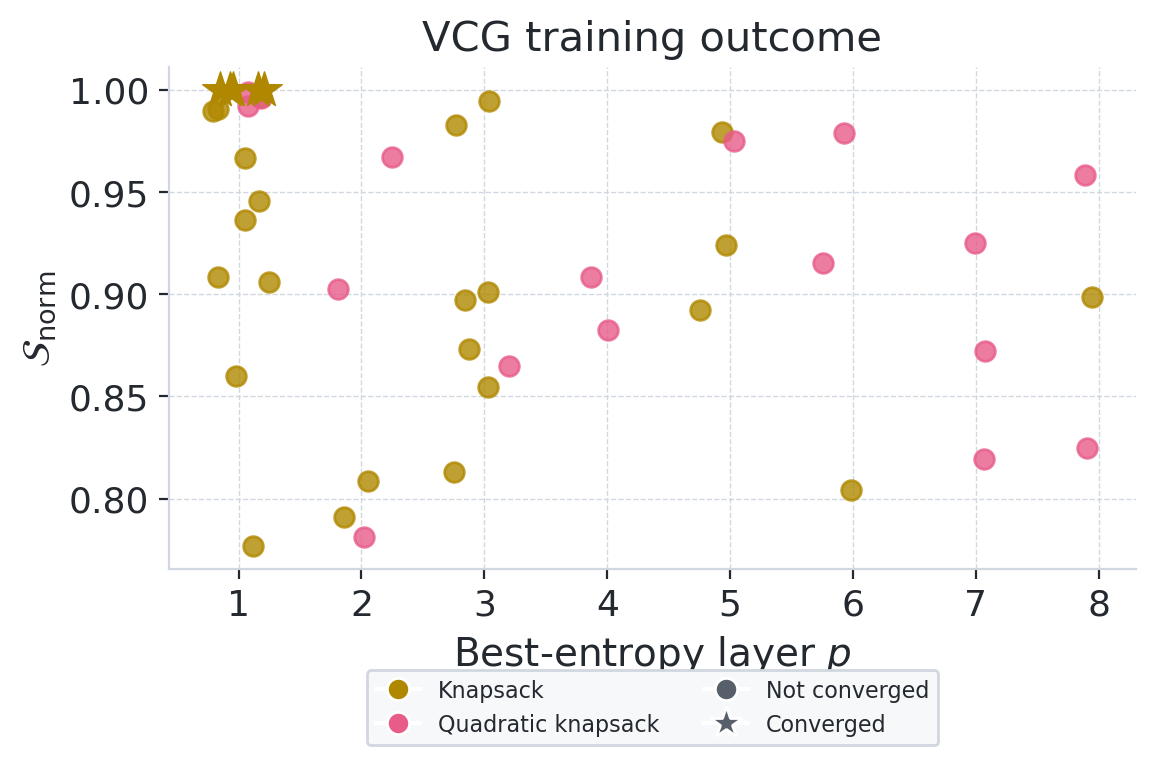}
  \caption{Normalized feasible entropy $\mathcal{S}_{\mathrm{norm}}$ versus best-entropy QAOA depth for the 46 QAOA-trained VCG gadgets (exact-preparation gadgets excluded). Stars indicate gadgets for which training terminated early upon jointly satisfying $P_{\mathcal{F}_k} \geq 0.999$ and $\mathcal{S}_{\mathrm{norm}} \geq 0.9999$.
  Circles denote gadgets that ran to the maximum depth $p_{\max} = 8$ without meeting both criteria, with the displayed layer showing the depth at which the highest $\mathcal{S}_{\mathrm{norm}}$ was recorded.}
  \label{fig:vcg_entropy_vs_depth}
\end{figure}

\begin{figure*}[t]
  \centering
  \includegraphics[width=0.95\linewidth]{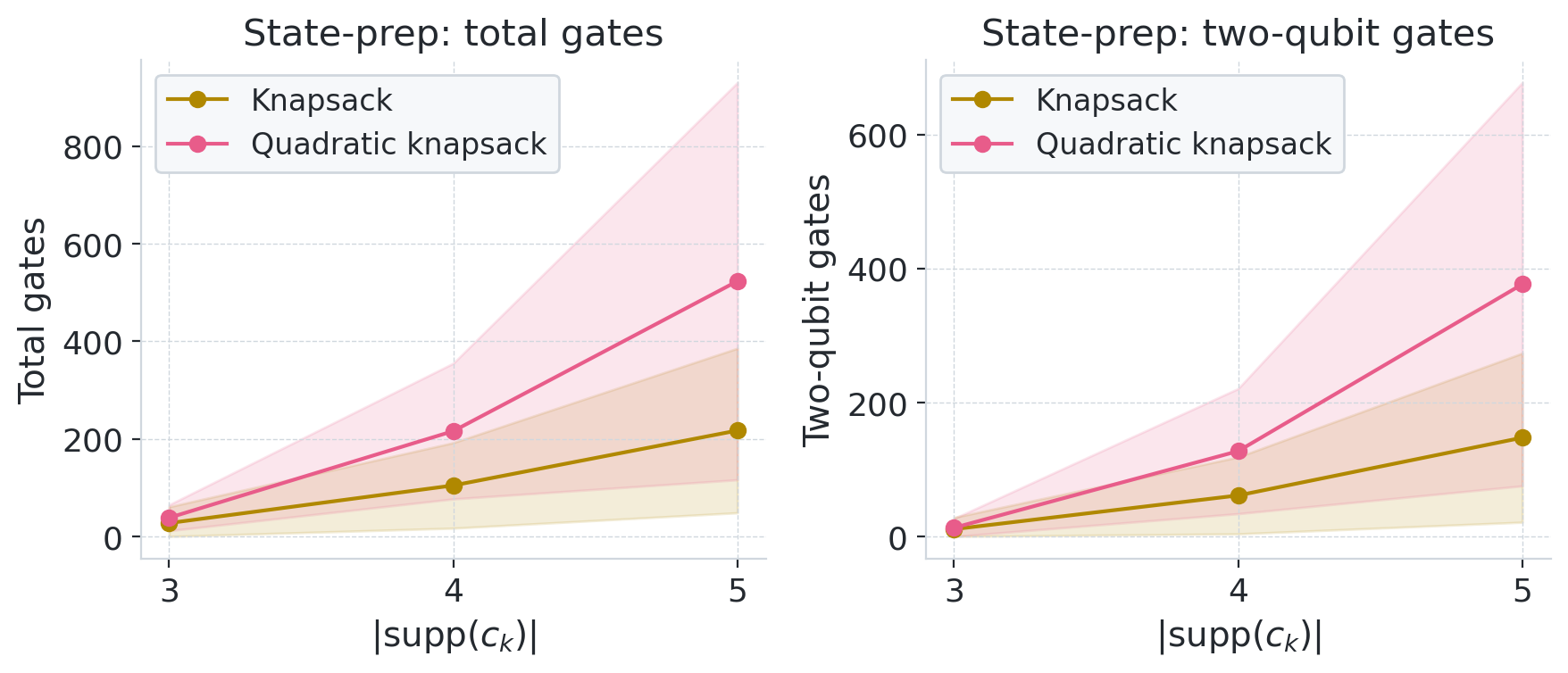}
  \caption{Circuit resources of the trained VCG state-preparation circuit vs.\
    problem size $n$, grouped by constraint family.
    \textbf{(a)}~Total gate count.
    \textbf{(b)}~Two-qubit gate count.
    Shaded regions show $\pm 1$ standard deviation.}
  \label{fig:vcg_circuit_resources}
\end{figure*}

In all cases, the resulting VCGs achieved $P_{\mathcal{F}_k} = 1$ showing that the approximated state contains only feasible states in the superposition.
Each VCG also achieved $\mathcal{S}_{\mathrm{norm}} > 0.7769$ with an average of 0.9186, which means that the states created also have near-uniform amplitudes across the feasible states.
Fig.~\ref{fig:vcg_entropy_by_type} shows a breakdown of $\mathcal{S}_{\mathrm{norm}}$ by constraint family across the number of variables in the support of the constraint $|\mathrm{supp}(c_k)|$.
As expected, $\mathcal{S}_{\mathrm{norm}}$ tends to decrease as $|\mathrm{supp}(c_k)|$ increases. Knapsack and quadratic-knapsack gadgets achieve similar mean $\mathcal{S}_{\mathrm{norm}}$ ($0.9204$ vs.\ $0.9155$).
However, a closer look reveals that $41$ out of $46$ of the trained VCGs had not converged, meaning that they trained up to 8 layers of ma-QAOA without reaching the $\mathcal{S}_{\mathrm{norm}} \ge  0.9999$ threshold.
In Fig.~\ref{fig:vcg_entropy_vs_depth}, we show the relationship between the number of layers at which the best $\mathcal{S}_{\mathrm{norm}}$ was achieved and the value of $\mathcal{S}_{\mathrm{norm}}$ itself.
We can see that the $5$ VCGs that did converge were those created for knapsack constraints and they each only took a single layer to do so.
All the other VCGs did not converge, but a majority of the knapsack VCGs achieved their best $\mathcal{S}_{\mathrm{norm}}$ within $3$ layers, while the quadratic knapsack VCGs typically required more layers to reach their best $\mathcal{S}_{\mathrm{norm}}$.
Rather than saving the VCG from the final training layer, we save the gadget from the layer achieving the best $\mathcal{S}_{\mathrm{norm}}$, as it requires fewer circuit resources and produces the most uniform amplitude distribution over the feasible states.
In Fig.~\ref{fig:vcg_circuit_resources}, we show the total gate count and two-qubit gate count of the resulting state-preparation circuits.

\subsection{PC-QAOA vs. PenaltyQAOA}\label{sec: pcqaoa-reselts}

In PC-QAOA, structural constraints are enforced via exact or VCG-based initial state preparation with a Grover mixer. Residual penalty constraints are added to the cost Hamiltonian with strength $\delta = 5 + 2|f_{\min}|$, where $f_{\min}$ is the unconstrained minimum of the QUBO. Although constraint-specific penalty weights may improve performance, we use a single shared penalty weight in both PC-QAOA and PenaltyQAOA for simplicity, ensuring a fair comparison. This heuristic scales $\delta$ with the objective magnitude and is sufficient to suppress infeasible solutions for the QUBOs used in our experiments.
PenaltyQAOA adds \emph{all} constraints as quadratic penalty terms with the same $\delta$ and uses a standard $X$-mixer.
Both methods sweep all layers $p = 1, \ldots, 5$. Shared optimization hyperparameters are listed in Table~\ref{tab:pcqaoa_hyperparams}.

\begin{table}[h]
\centering
\caption{Shared optimization hyperparameters for PC-QAOA and PenaltyQAOA.}
\label{tab:pcqaoa_hyperparams}
\begin{tabular}{lc}
\hline
\textbf{Parameter} & \textbf{Value} \\
\hline
Maximum layers ($p_{\max}$)   & 5  \\
Restarts per layer            & 10 \\
Optimization steps            & 50 \\
Learning rate (Adam)          & 0.01 \\
Measurement shots             & 10,000 \\
Penalty weight $\delta$       & $5+2|f_{\min}|$ \\
\hline
\end{tabular}
\end{table}

\begin{figure*}[t]
  \centering
  \begin{subfigure}[b]{0.48\linewidth}
    \includegraphics[width=\linewidth]{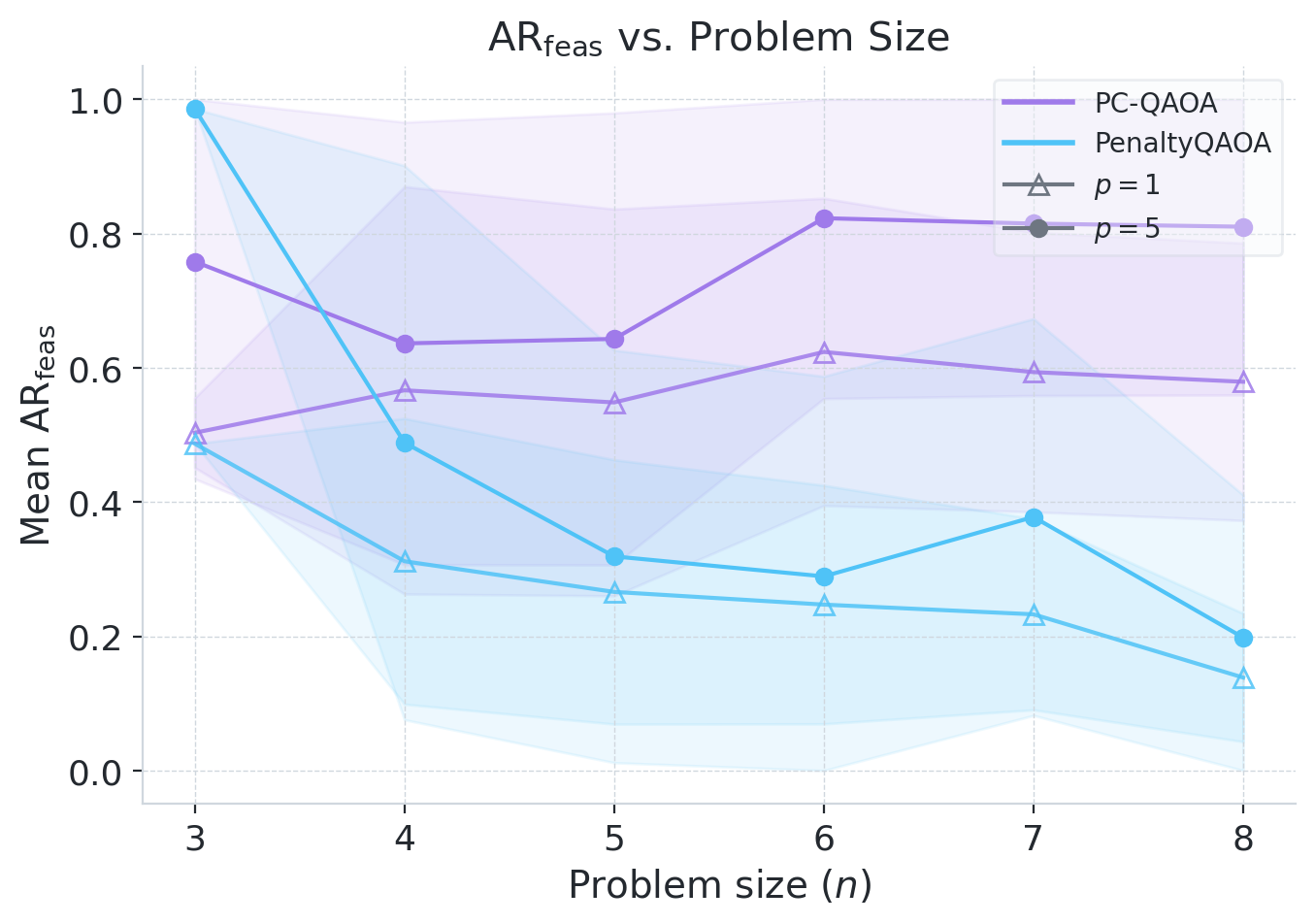}
    \caption{PC-QAOA vs. PenaltyQAOA across all problems tested.}
    \label{fig:ar_feas_vs_nx_a}
  \end{subfigure}
  \hfill
  \begin{subfigure}[b]{0.48\linewidth}
    \includegraphics[width=\linewidth]{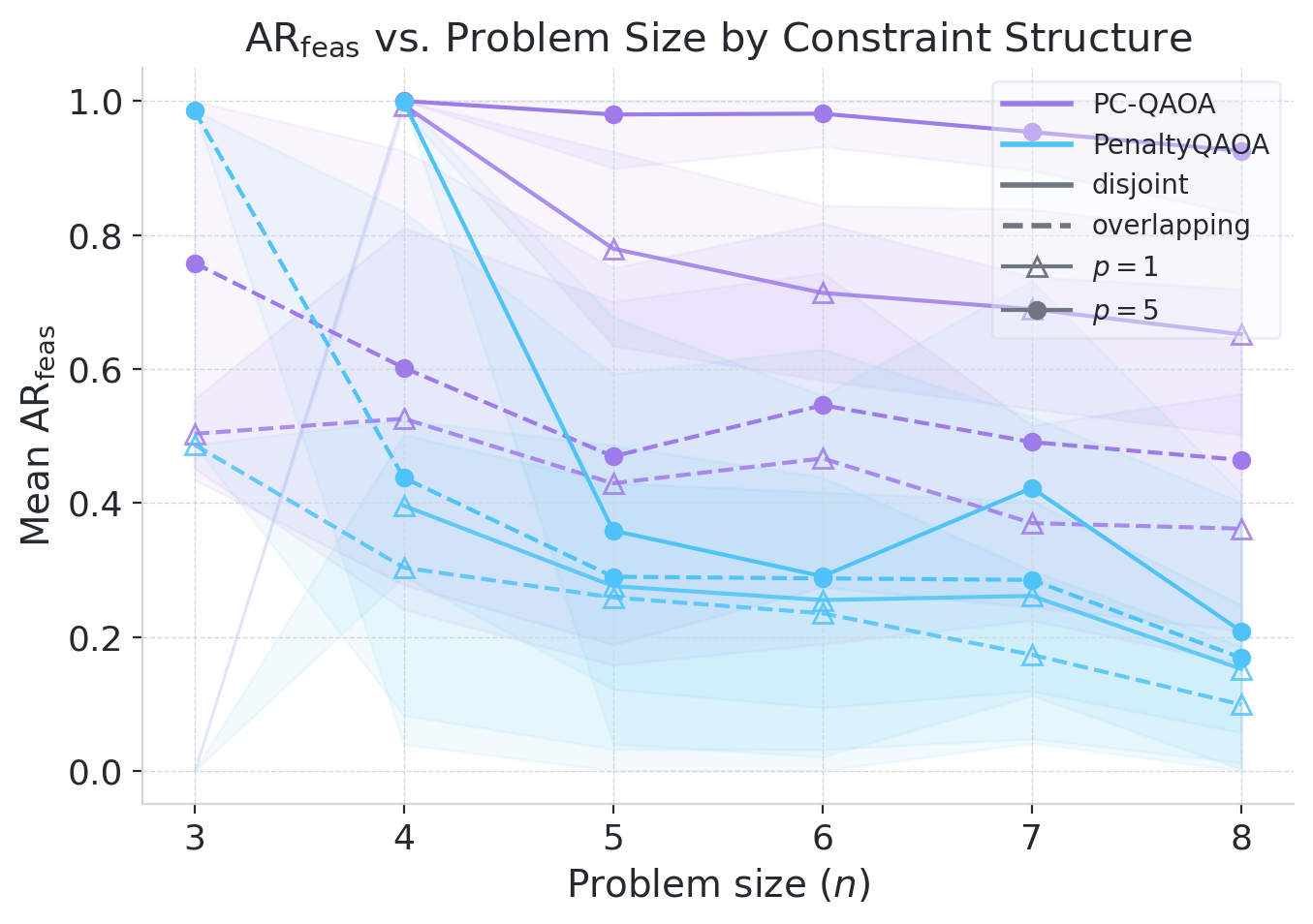}
    \caption{Both methods split by whether the problem contains overlapping constraints or only disjoint constraints.}
    \label{fig:ar_feas_vs_nx_b}
  \end{subfigure}
  \caption{Mean $\text{AR}_{\text{feas}}$ vs. problem size $n$.
    Instances where $P(\text{feas})=0$ are assigned $\text{AR}_{\text{feas}}=0$.
    Shaded regions show $\pm 1$ standard deviation.}
  \label{fig:ar_feas_vs_nx}
\end{figure*}

The first metric of interest is the feasible approximation ratio
\begin{equation*}
  \text{AR}_{\text{feas}}
    = \frac{\bar{f}_{\text{feas}} - f_{\max}^{\mathcal{F}}}{f_{\min}^{\mathcal{F}} - f_{\max}^{\mathcal{F}}},
  \qquad
  \bar{f}_{\text{feas}}
    = \frac{\sum_{x \in \mathcal{F}} f(x)\,p(x)}{P(\text{feas})},
\end{equation*}
where $p(x)=|\langle x|\psi\rangle|^2$, $\mathcal{F}$ is the feasible set, and $f_{\min}^{\mathcal{F}}$ and $f_{\max}^{\mathcal{F}}$ are the minimum and maximum objective values over $\mathcal{F}$.
Unlike the raw approximation ratio, $\text{AR}_{\text{feas}}$ is unaffected by penalty inflation and gives a fair cross-method comparison of solution quality.
As shown in Fig.~\ref{fig:ar_feas_vs_nx_a}, PC-QAOA maintains higher solution quality across all sizes at both $p=1$ (mean $0.588$ vs. $0.219$) and $p=5$ (mean $0.783$ vs. $0.305$), while PenaltyQAOA degrades more steeply as $n$ grows.
Fig.~\ref{fig:ar_feas_vs_nx_b} shows mean $\text{AR}_{\text{feas}}$ vs. problem size $n$ split by whether the problem contains overlapping constraints or only disjoint constraints.
While PC-QAOA outperforms PenaltyQAOA in both cases, the presence of overlapping constraints results in a lower $\text{AR}_{\text{feas}}$ for PC-QAOA.
This degradation is expected, as overlapping constraints require penalization.
Encoding structural constraints into the circuit substantially improves the approximation ratio across all problem sizes.

\begin{figure*}[t]
  \centering
  \begin{subfigure}[b]{0.48\linewidth}
    \includegraphics[width=\linewidth]{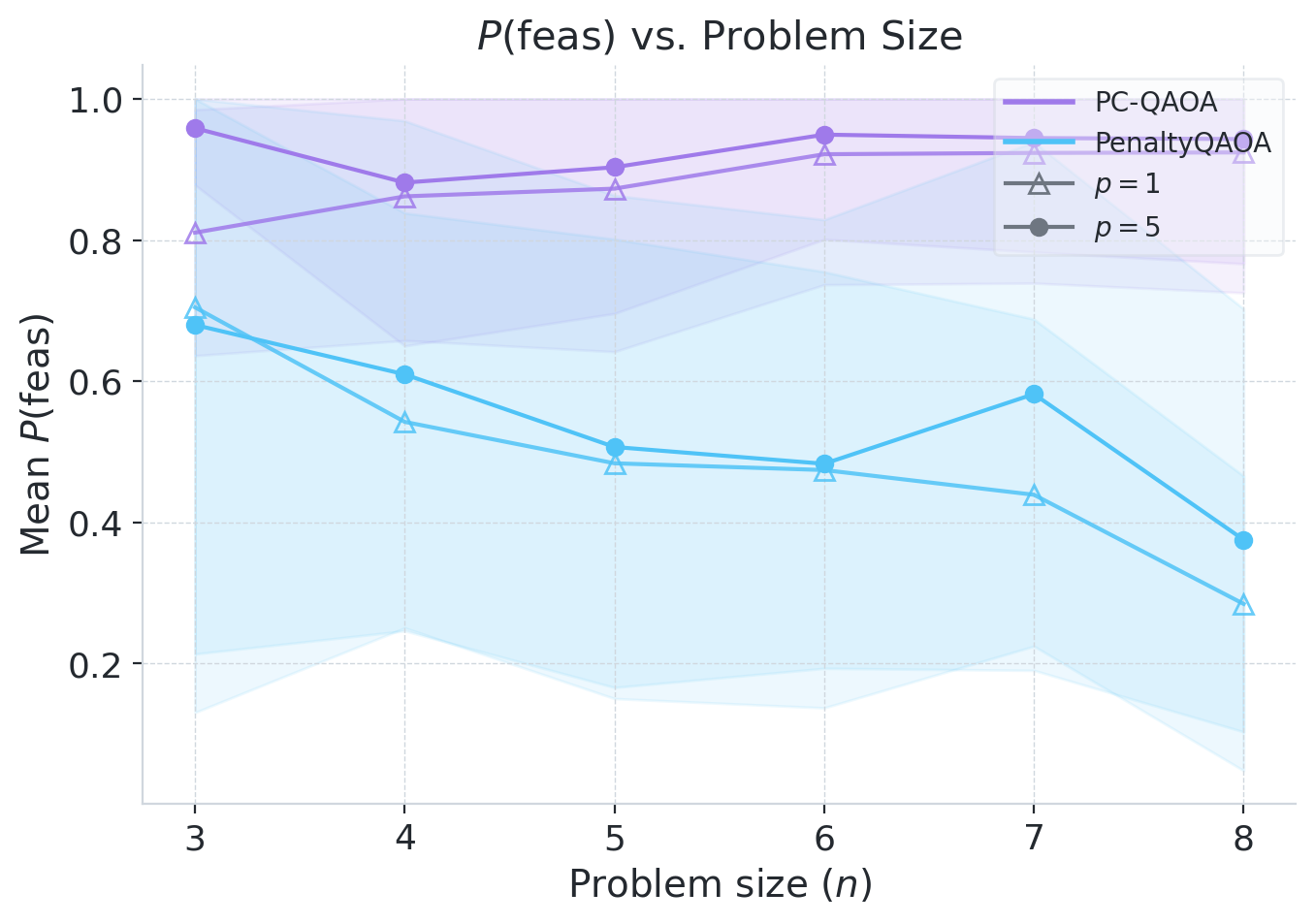}
    \caption{PC-QAOA vs. PenaltyQAOA, all instances combined.
      Filled circles: $p=5$; open triangles: $p=1$.}
    \label{fig:pcqaoa_vs_penalty_pfeas_a}
  \end{subfigure}
  \hfill
  \begin{subfigure}[b]{0.48\linewidth}
    \includegraphics[width=\linewidth]{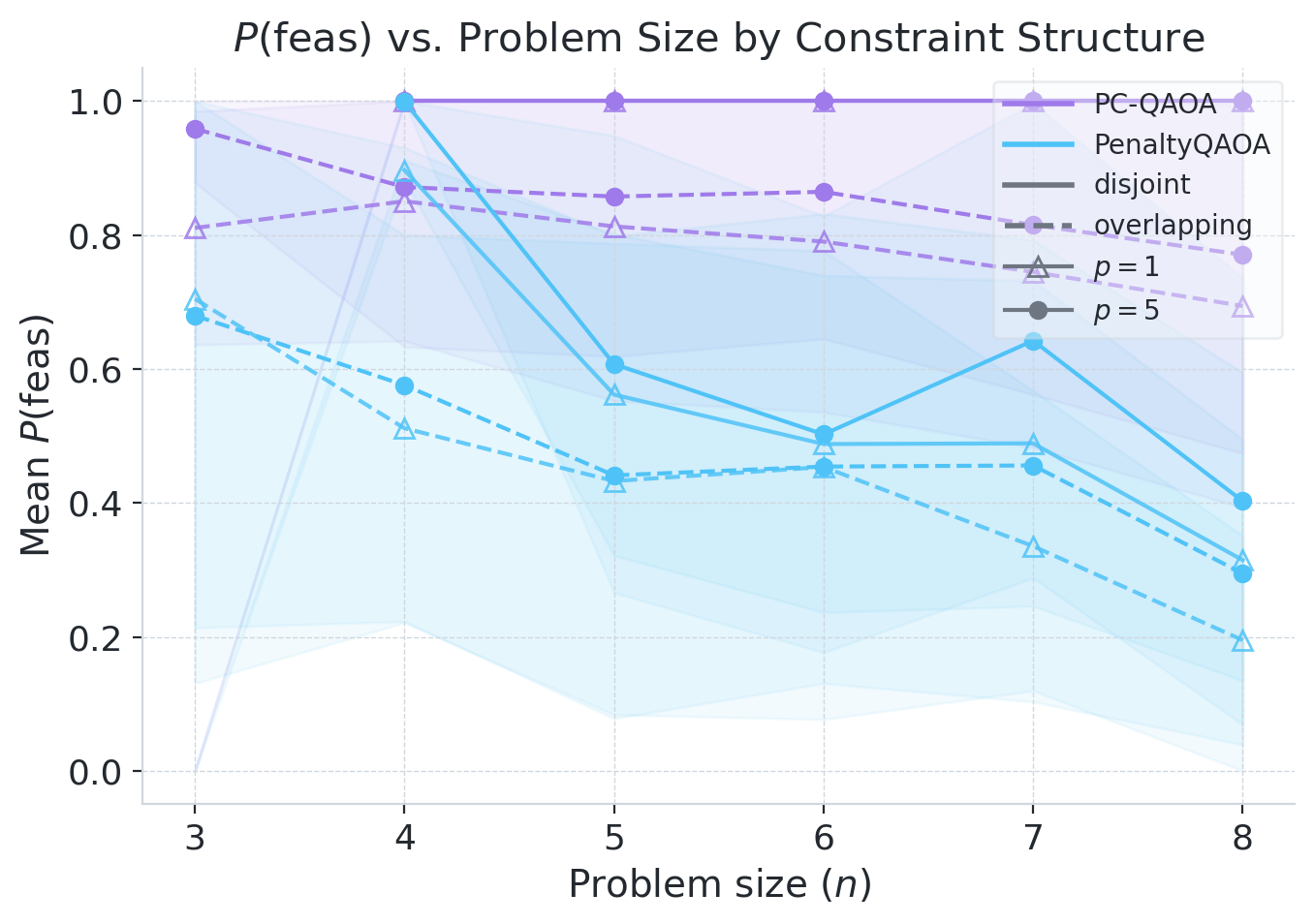}
    \caption{Both methods split by whether the problem contains overlapping constraints or only disjoint constraints.}
    \label{fig:pcqaoa_vs_penalty_pfeas_b}
  \end{subfigure}
  \caption{Mean $P(\text{feas})$ vs. $n$.
    Shaded regions show $\pm 1$ standard deviation.}
  \label{fig:pcqaoa_vs_penalty_pfeas}
\end{figure*}

The second metric we use is the probability of sampling a feasible solution, $P(\text{feas})$.
Fig.~\ref{fig:pcqaoa_vs_penalty_pfeas_a} shows $P(\text{feas})$ as a function of problem size $n$.
We can see that PC-QAOA maintains a high $P(\text{feas})$ as $n$ increases, while PenaltyQAOA's $P(\text{feas})$ degrades more steeply with $n$.
Specifically, PC-QAOA achieves mean $P(\text{feas}) = 0.912$ at $p=1$, rising to $0.936$ at $p=5$, while PenaltyQAOA reaches only $0.420$ at $p=1$ and $0.491$ at $p=5$.
Since PC-QAOA starts in a feasible subspace of the full search space, with mixers designed to preserve this subspace, it is expected to retain this high $P(\text{feas})$.
As shown in Fig.~\ref{fig:pcqaoa_vs_penalty_pfeas_b}, the disjoint constraint problems retain a $P(\text{feas}) = 1$.
However, in the presence of penalized constraints, the initial state generally contains states that are infeasible with respect to those constraints, and the mixer no longer preserves their feasible subspace. Consequently, $P(\mathrm{feas})$ is expected to decrease for problems with overlapping constraints compared to disjoint constraints.

\begin{figure*}[t]
  \centering
  \begin{subfigure}[b]{0.48\linewidth}
    \includegraphics[width=\linewidth]{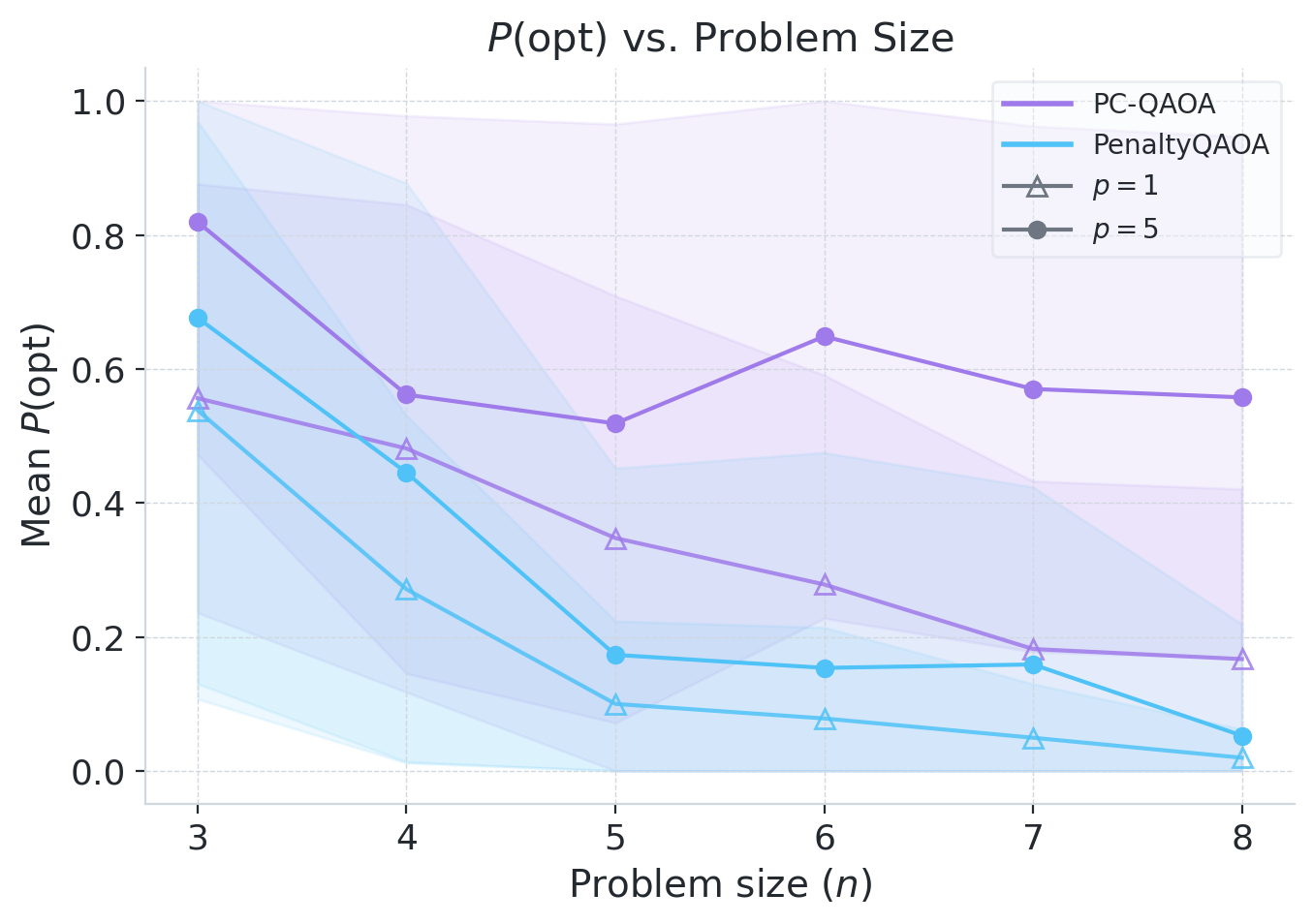}
    \caption{PC-QAOA vs. PenaltyQAOA, all instances combined.}
    \label{fig:pcqaoa_p_opt_a}
  \end{subfigure}
  \hfill
  \begin{subfigure}[b]{0.48\linewidth}
    \includegraphics[width=\linewidth]{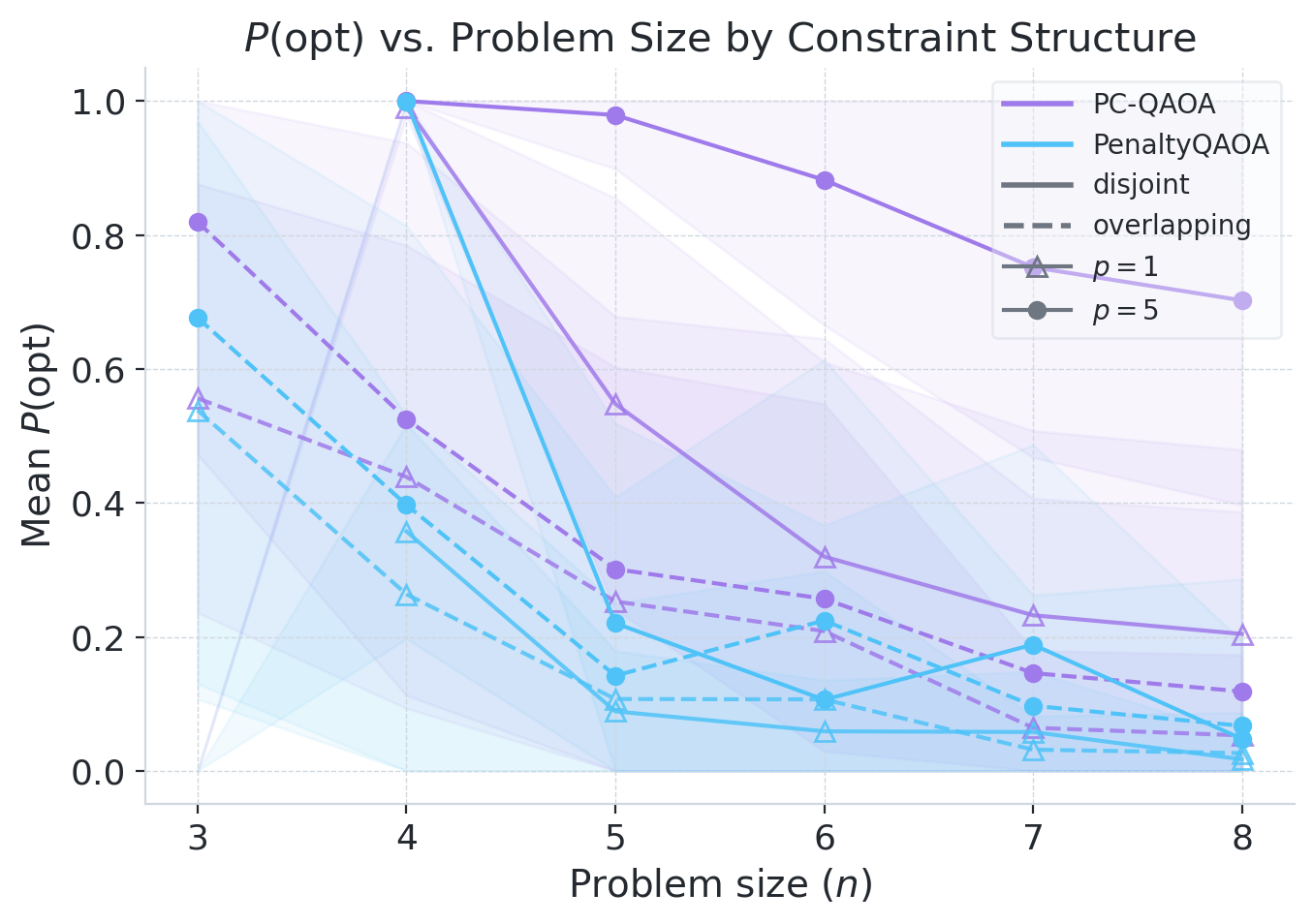}
    \caption{Both methods split by whether the problem contains overlapping constraints or only disjoint constraints.}
    \label{fig:pcqaoa_p_opt_b}
  \end{subfigure}
  \caption{Mean $P(\text{opt})$ vs. $n$.
    Shaded regions show $\pm 1$ standard deviation.}
  \label{fig:pcqaoa_p_opt}
\end{figure*}

The last metric we use is the probability of sampling the optimal solution, $P(\text{opt})$.
Fig.~\ref{fig:pcqaoa_p_opt_a} shows $P(\text{opt})$ vs. $n$.
PC-QAOA achieves mean $P(\text{opt}) = 0.242$ at $p=1$ vs. $0.077$ for PenaltyQAOA, a $215\%$ relative improvement, improving further to $0.579$ vs. $0.156$ at $p=5$.
Both methods see $P(\text{opt})$ decline with problem size $n$ as the feasible search space grows.
Similar to the previous two metrics, Fig.~\ref{fig:pcqaoa_p_opt_b} shows that problems with overlapping constraints result in a lower $P(\text{opt})$ than disjoint problems for PC-QAOA, especially at $p=5$.
PenaltyQAOA, while displaying a lower $P(\text{opt})$ than PC-QAOA, does not show a considerable difference in the presence of overlapping constraints.

\begin{figure*}[t]
  \includegraphics[width=0.75\linewidth]{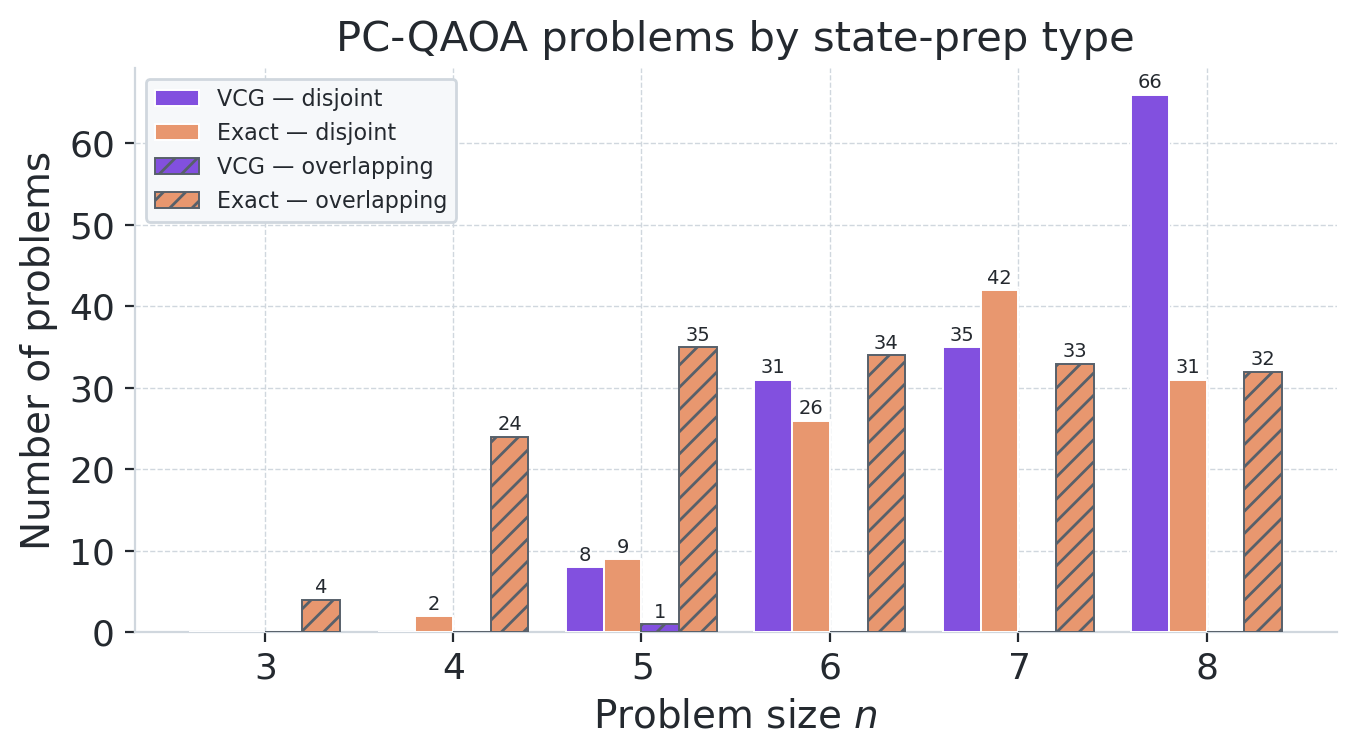}
  \caption{Instance counts for PC-QAOA split by whether the instance includes at least one of the 46 constraints implemented with a VCG-based state preparation or only constraints with exact Dicke-state preparations, and by whether the problem contains overlapping constraints or only disjoint constraints.}
  \label{fig:vcg_vs_exact_counts}
\end{figure*}

Fig.~\ref{fig:vcg_vs_exact_counts} shows the distribution of problem instances by whether they include at least one of the $46$ constraints implemented with a VCG-based state preparation or only constraints with exact Dicke-state preparations, and by whether the problem contains overlapping constraints or only disjoint constraints.
VCG instances are nearly all disjoint ($140$ of $141$), while the exact gadget group contains a large proportion of overlapping instances ($162$ of $272$).
There are no VCG instances at $n=3$ or $n=4$, since they require at least two structural constraints and the smallest VCG has support on three variables.

\begin{figure*}[t]
  \centering
  \begin{subfigure}[b]{0.32\linewidth}
    \includegraphics[width=\linewidth]{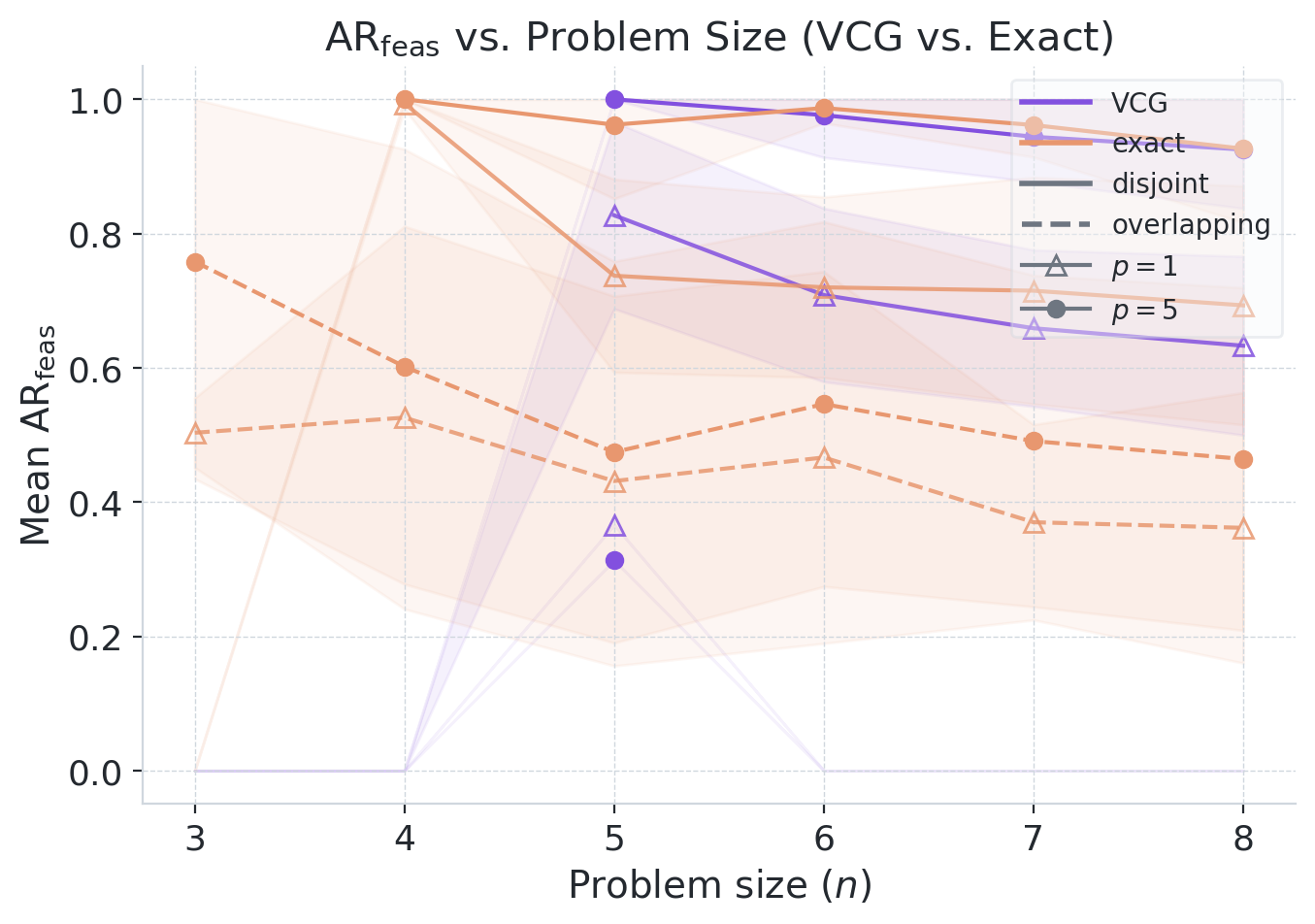}
    \caption{$\text{AR}_{\text{feas}}$ vs. $n$.}
    \label{fig:vcg_vs_exact_ar}
  \end{subfigure}
  \hfill
  \begin{subfigure}[b]{0.32\linewidth}
    \includegraphics[width=\linewidth]{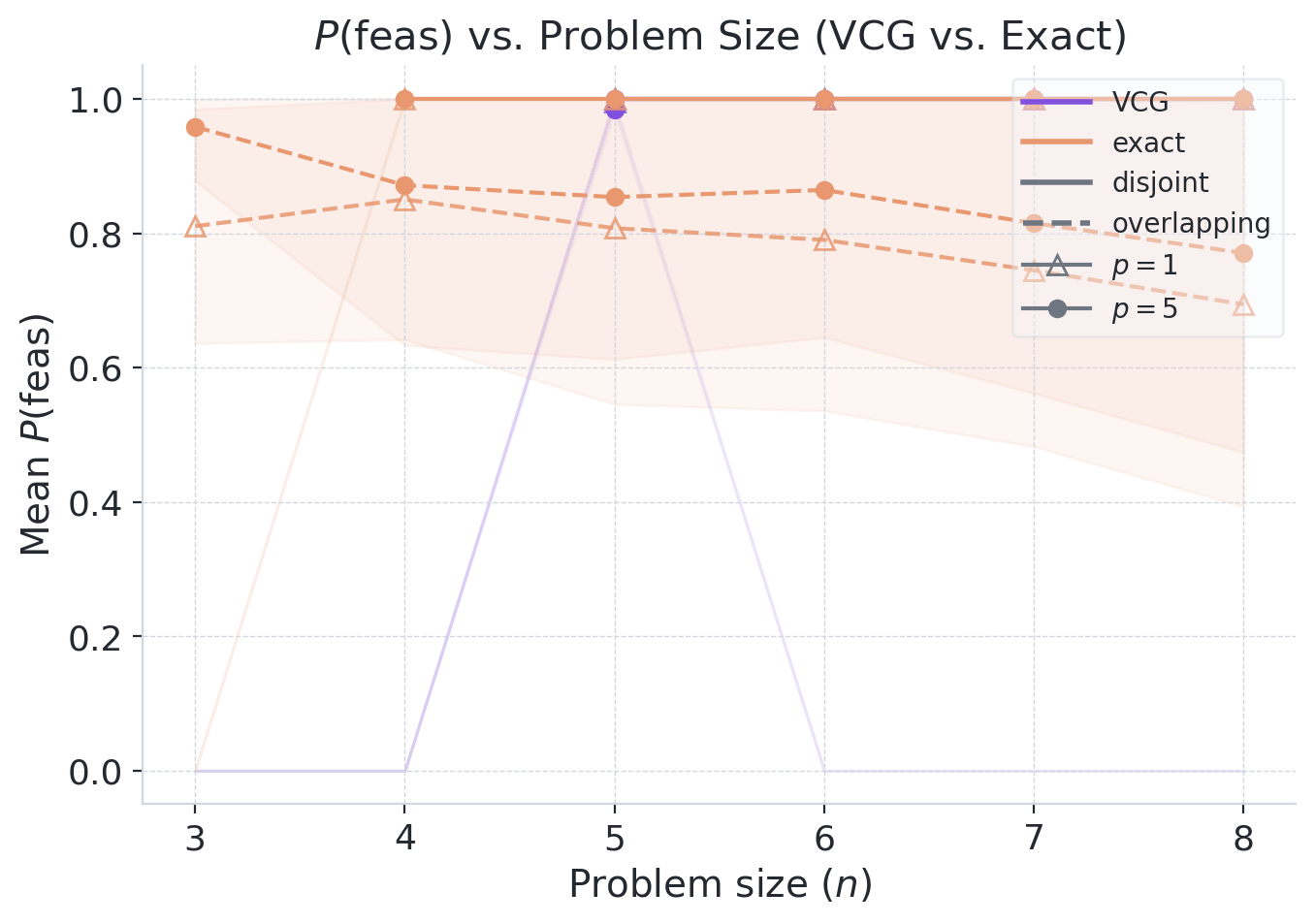}
    \caption{$P(\text{feas})$ vs. $n$.}
    \label{fig:vcg_vs_exact_pfeas}
  \end{subfigure}
  \hfill
  \begin{subfigure}[b]{0.32\linewidth}
    \includegraphics[width=\linewidth]{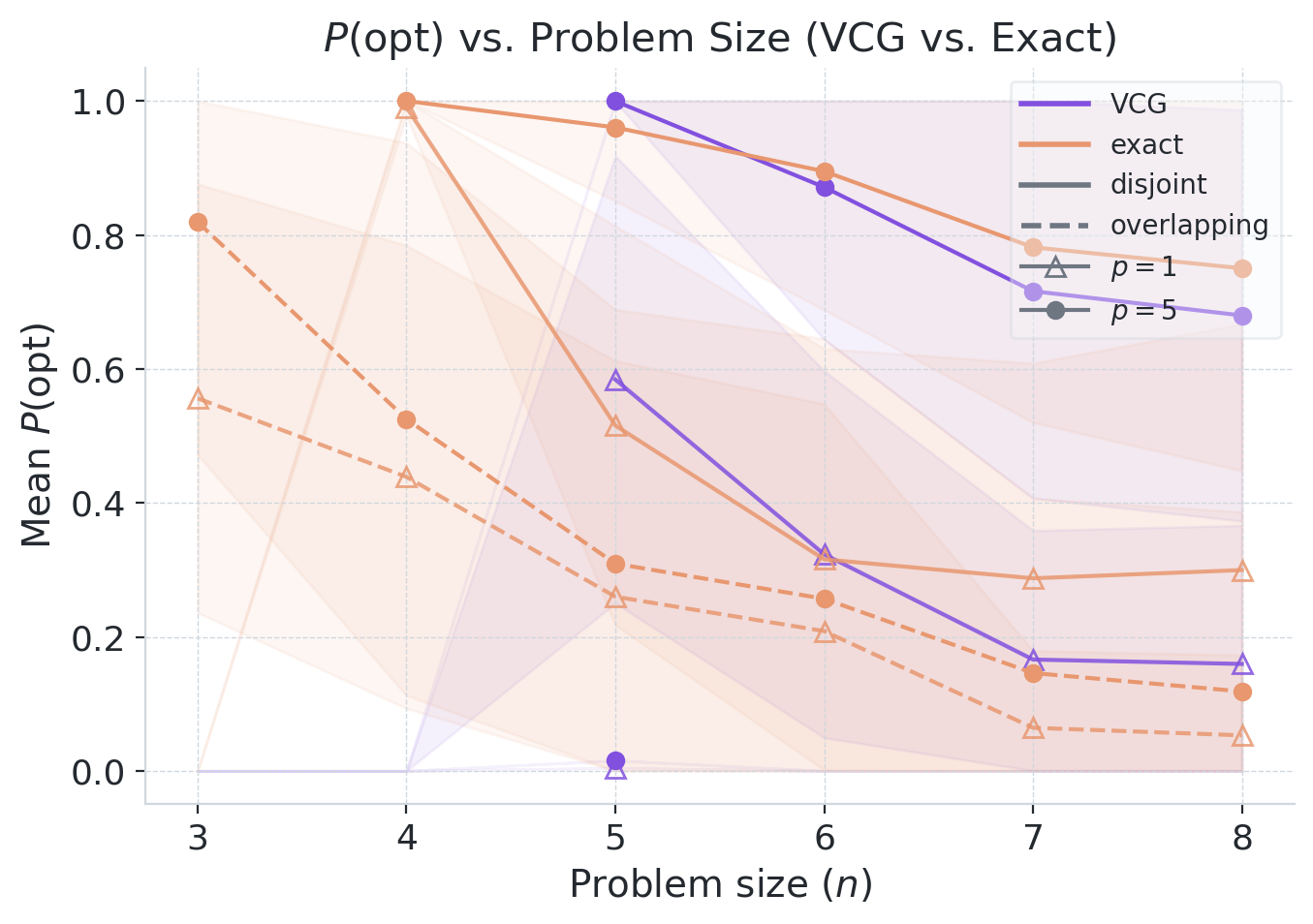}
    \caption{$P(\text{opt})$ vs. $n$.}
    \label{fig:vcg_vs_exact_popt}
  \end{subfigure}
  \caption{PC-QAOA performance split by whether the instance includes at least one of the 46 constraints implemented with a VCG-based state preparation or only constraints with exact Dicke-state preparations, and by whether the problem contains overlapping constraints or only disjoint constraints. Solid lines show disjoint instances; the single overlapping VCG instance appears as an isolated marker. Filled markers show $p=5$; open markers show $p=1$. Shaded regions show $\pm 1$ standard deviation.}
  \label{fig:vcg_vs_exact}
\end{figure*}

Fig.~\ref{fig:vcg_vs_exact} compares $\mathrm{AR}_{\text{feas}}$, $P(\text{feas})$, and $P(\text{opt})$ for these two groups, restricted to problems with only disjoint constraints to control for the imbalance.
Among disjoint instances, both groups achieve $P(\text{feas}) = 1.000$ at all depths.
Exact preparation achieves slightly higher $\mathrm{AR}_{\text{feas}}$ ($0.958$ vs.\ $0.945$ at $p=5$) and $P(\text{opt})$ ($0.818$ vs.\ $0.750$ at $p=5$), consistent with VCGs producing near-uniform but not perfectly uniform superpositions ($\mathcal{S}_{\mathrm{norm}} \approx 0.92$ on average vs.\ $1.0$ for exact).
The gap is larger at $p=1$ and narrows with depth, suggesting the outer QAOA layers partially compensate for imperfect VCG state preparation.

%

\begin{figure*}[t]
  \centering
  \includegraphics[width=0.95\linewidth]{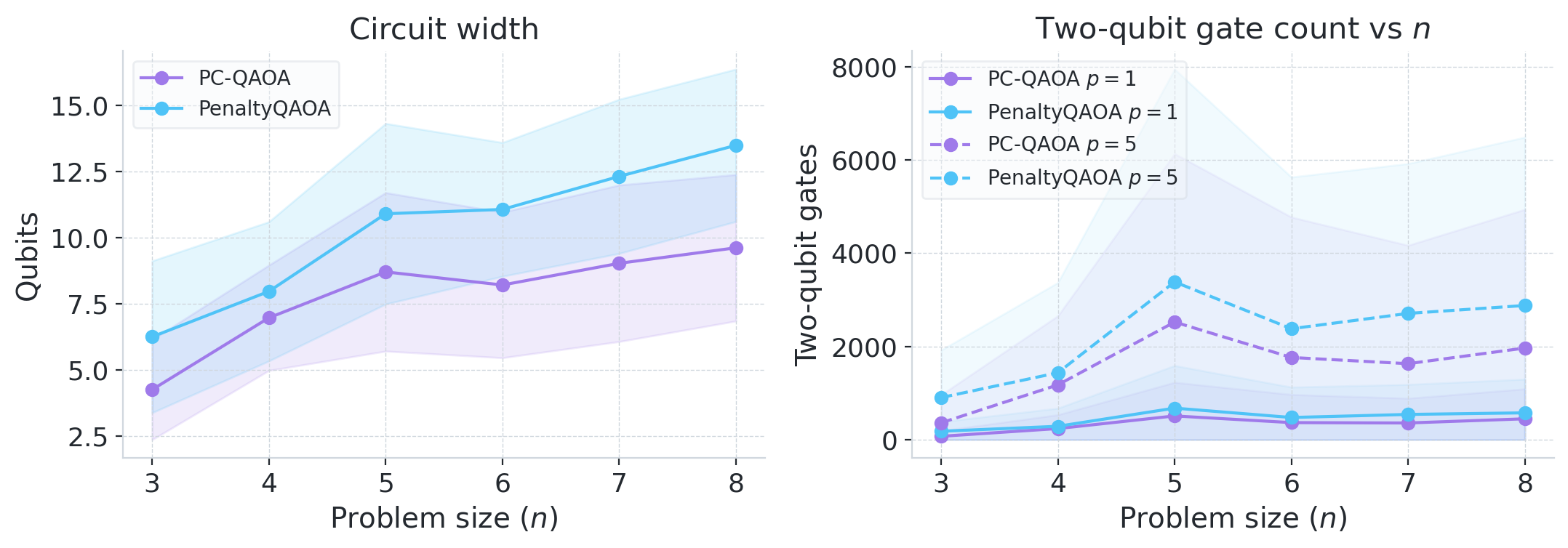}
  \caption{Circuit resource comparison for PC-QAOA and PenaltyQAOA.
    \textbf{(a)}~Total qubit count vs. $n$:
    PC-QAOA requires no slack qubits for structurally enforced constraints,
    while PenaltyQAOA introduces one slack qubit per inequality constraint.
    \textbf{(b)}~Total two-qubit gate count (state preparation plus $p$ cost
    and mixer layers) vs. $n$, shown at $p=1$ (solid) and $p=5$ (dashed).
    Shaded regions show $\pm 1$ standard deviation.}
  \label{fig:circuit_resources}
\end{figure*}

On near-term devices, circuit depth is the primary bottleneck imposed by decoherence. PC-QAOA achieves the feasibility threshold at $p=1$ for 85.5\% of instances, requiring far shallower circuits than PenaltyQAOA, which typically needs $p=3$--$5$ for the same outcome.
Fig.~\ref{fig:circuit_resources} quantifies the circuit resource difference between PC-QAOA and PenaltyQAOA.
Because structural constraints are enforced through the initial state and mixer rather than penalty terms, PC-QAOA requires no slack qubits for those constraints.
This gives a mean qubit count of $8.8$ vs.\ $11.8$ for PenaltyQAOA across all instances ($7.0$ vs.\ $11.4$ for disjoint instances).
PC-QAOA does, however, incur additional state-preparation overhead from VCG and Dicke-state circuits.
PenaltyQAOA instead initializes each qubit with a Hadamard gate, requiring no two-qubit state-preparation gates.
Despite this overhead, PC-QAOA achieves a net reduction in total two-qubit gate count, requiring $1.35\times$ fewer two-qubit gates at $p=1$ and $1.45\times$ fewer at $p=5$ across all instances, since the removal of penalty terms from the cost Hamiltonian reduces the number of two-qubit gates required per QAOA layer.
For disjoint instances, where all constraints are structurally enforced and no slack qubits are needed, the advantage grows to $1.64\times$ fewer two-qubit gates at $p=1$ and $2.12\times$ fewer at $p=5$.

\section{Conclusion and Future Work}\label{sec: Discussion}
In this work we introduce PC-QAOA, a framework that partitions the constraints of a combinatorial optimization problem into two sets: those that can be efficiently enforced via structural encoding in the initial state and mixer, and those that are handled via penalty terms in the cost Hamiltonian.
The key theoretical result (Proposition~\ref{prop:composition}) is that gadgets for constraints with pairwise disjoint supports compose exactly without interference, so the structural encoding can be applied to multiple constraints simultaneously without additional overhead.
 
 We show classes of constraints that are natural candidates for structural enforcement. To handle constraint families lacking exact constructions, we introduced VCGs: parameterized circuits trained offline via a two-stage multi-angle QAOA procedure.
 All 60 VCGs evaluated across knapsack and quadratic-knapsack families ($|\mathrm{supp}(c_k)| \in \{3,4,5\}$) achieve $P_{\mathcal{F}_k} = 1$ and mean $\mathcal{S}_{\mathrm{norm}} = 0.9186$ within at most 8 training layers. Among disjoint instances, exact Dicke-state preparation modestly outperforms VCG preparation ($\mathrm{AR}_{\mathrm{feas}}$ $0.958$ vs.\ $0.945$, $P(\mathrm{opt})$ $0.818$ vs.\ $0.750$ at $p=5$), consistent with VCGs achieving near- but not perfectly uniform feasible superpositions.

Evaluated on 413 completed instances spanning seven constraint families and $n \in \{3,\ldots,8\}$, PC-QAOA achieves mean $P(\text{feas}) = 0.912$ vs. $0.420$ at $p=1$ and mean $P(\text{opt}) = 0.242$ vs. $0.077$ compared to PenaltyQAOA.
PC-QAOA meets the $P(\text{feas}) \geq 0.75$ threshold at $p=1$ in 85.5\% of completed instances vs. 15.2\%, and 65.5\% of PenaltyQAOA instances fail to converge within $p_{\max}=5$ layers vs. 9.4\% for PC-QAOA.
Structural enforcement also eliminates the slack qubits required by penalty methods and reduces two-qubit gate count, yielding direct circuit resource savings that are important for near-term hardware.

While these results demonstrate the promise of PC-QAOA, they also highlight limitations and avenues for future work.
The effectiveness of PC-QAOA depends on the availability of a useful partition of the constraint set into its structural and energetically enforced components.
When constraints have large supports, dense overlap, or little exploitable symmetry, only a small fraction of the instance may admit a practical structural encoding, in which case PC-QAOA moves closer to penalty-based approaches.
Also, although PC-QAOA reduces the reliance on penalty terms, it does not remove this reliance altogether.
Energetically enforced constraints still require suitable penalty weights, and solution quality may depend on this choice even within the reduced feasible region.


A key limitation of PC-QAOA's structural encoding component is its scalability to large-support constraints.
A natural direction for addressing this is classical preprocessing to reduce the effective support size before quantum encoding.
Cover inequalities~\cite{Balas1975, Gu1998} provide one such tool.
For example, with knapsack-type constraints, minimal cover reductions can replace a high-arity weighted inequality with a tighter cardinality constraint of smaller support, directly reducing gadget complexity.
Extended covers and lifting procedures~\cite{Gu1998} yield even tighter polyhedral approximations.
More broadly, bound propagation, symmetry detection, and constraint aggregation techniques from integer programming could be combined with gadget-based encoding to extend the approach to problem instances with denser or higher-arity constraints.

Another direction for future work is to apply PC-QAOA to a broader set of industry-relevant constrained binary optimization problems.
While this paper focuses on benchmark constraint families, many real-world applications involve richer constraints: vehicle routing with capacity and time windows, workforce scheduling, and supply-chain network design, where penalty-based encoding becomes cumbersome and problem-specific mixer design is impractical.

\section*{Acknowledgments}
A.D., A.D.R., and R.H. acknowledge NSF CNS-2244512.
A.W., R.H., and J.O. acknowledge NSF CCF 2210063 and FA2385-25-C-B012.

\section*{Code and Data Availability}
\label{sec:codeAndDataAvail}
The code and data for this research can be found at \url{https://github.com/Vilcius/constraint_gadgets}.

\section*{Use of AI Assistance}

Large language models (Claude, Anthropic) were used during the preparation of this work to assist with manuscript editing (e.g., restructuring text, improving clarity, and consistency checks) and to support aspects of software development, including code prototyping and debugging.

AI tools were not used to generate scientific hypotheses, design experiments, or interpret results.
All algorithms, experimental design choices, and conclusions were developed by the authors.

All AI-assisted code and text were reviewed, validated, and, where applicable, independently reproduced by the authors to ensure correctness and reliability.
The final experimental pipeline and results are fully reproducible from the provided codebase.
\renewcommand\refname{References Cited}
\bibliography{references}
\bibliographystyle{IEEEtran}

\appendix

\section{Sequential State-Preparation Gadgets for Overlapping Constraints}\label{app:overlapping_stateprep}

When multiple constraints have overlapping support, it is natural to ask whether sequential application of state-preparation unitaries yields the uniform superposition over the intersection of feasible regions.
We show that, in general, it does not.

Let $\mathcal{F}_0, \mathcal{F}_1 \subseteq \{0,1\}^n$ denote the feasible sets of two classical constraints, possibly with overlapping supports.
Let $U_0$ and $U_1$ be unitaries satisfying
\begin{align*}
U_0 |0^n\rangle &= \frac{1}{\sqrt{|\mathcal{F}_0|}} \sum_{x \in \mathcal{F}_0} |x\rangle, \\
U_1 |0^n\rangle &= \frac{1}{\sqrt{|\mathcal{F}_1|}} \sum_{x \in \mathcal{F}_1} |x\rangle.
\end{align*}
These conditions specify only the action of $U_j$ on $|0^n\rangle$, their action on the remaining computational basis states is determined by unitarity but otherwise unconstrained.

After the first preparation,
\begin{equation*}
|\psi_1\rangle
=
\frac{1}{\sqrt{|\mathcal{F}_0|}}
\sum_{x \in \mathcal{F}_0} |x\rangle.
\end{equation*}
Applying $U_1$ yields
\begin{equation*}
|\psi_2\rangle
=
U_1 |\psi_1\rangle
=
\frac{1}{\sqrt{|\mathcal{F}_0|}}
\sum_{x \in \mathcal{F}_0} U_1 |x\rangle.
\end{equation*}
Expanding in the computational basis, where $\alpha_{y|x} = \langle y|U_1|x\rangle \in \mathbb{C}$ are the matrix elements of $U_1$,
\begin{equation*}
U_1 |x\rangle
=
\sum_{y \in \{0,1\}^n}
\alpha_{y|x} |y\rangle,
\end{equation*}
so that
\begin{equation*}
|\psi_2\rangle
=
\sum_{y \in \{0,1\}^n}
\left(
\frac{1}{\sqrt{|\mathcal{F}_0|}}
\sum_{x \in \mathcal{F}_0}
\alpha_{y|x}
\right)
|y\rangle.
\end{equation*}
For $|\psi_2\rangle$ to be supported only on $\mathcal{F}_0 \cap \mathcal{F}_1$, it is necessary that
\begin{equation*}
\sum_{x \in \mathcal{F}_0} \alpha_{y|x} = 0
\quad
\text{for all } y \notin \mathcal{F}_0 \cap \mathcal{F}_1.
\end{equation*}
Because the coefficients $\alpha_{y|x}$ are subject only to the column-normalization constraint $\sum_y |\alpha_{y|x}|^2 = 1$ for each $x$, such cancellation holds only for specially constructed $U_1$.
In particular, unless $U_1$ preserves the entire subspace $\mathrm{span}\{|x\rangle : x \in \mathcal{F}_0\}$, the coefficients $\alpha_{y|x}$ will generally be nonzero for $y \notin \mathcal{F}_0$, contributing probability mass outside $\mathcal{F}_0 \cap \mathcal{F}_1$.
Thus, sequential state preparation 
unitaries do not, in general, implement intersection of feasible regions when supports overlap.

A minimal counterexample is given by two overlapping cardinality constraints on three qubits
\begin{align*}
x_1 + x_2 &= 1 \\
x_2 + x_3 &= 1.
\end{align*}
The feasible sets are
\begin{align*}
  \mathcal{F}_0 &= \{100,010,101,011\},
  \mathcal{F}_1 &= \{001,010,101,110\},
\end{align*}
with intersection $\mathcal{F}_0 \cap \mathcal{F}_1 = \{010, 101\}$.
Applying $U_{D_{12,1}}$ to qubits $1,2$ (qubit $3$ initialized to $|0\rangle$) gives $\frac{1}{\sqrt{2}}(|100\rangle + |010\rangle)$.
Applying $U_{D_{23,1}}$ to qubits $2,3$, and using $U_{D_{23,1}}|00\rangle_{23} = \frac{1}{\sqrt{2}}(|10\rangle+|01\rangle)_{23}$, the $|100\rangle = |1\rangle_1|00\rangle_{23}$ component yields
\begin{equation*}
\frac{1}{\sqrt{2}}\bigl(|110\rangle + |101\rangle\bigr),
\end{equation*}
while the $|010\rangle = |0\rangle_1|10\rangle_{23}$ component maps to $|0\rangle_1|\phi\rangle_{23}$ for some $|\phi\rangle_{23}$ orthogonal to $U_{D_{23,1}}|00\rangle_{23}$ by unitarity.
The string $110$ satisfies the second constraint ($x_2+x_3=1$) but violates the first ($x_1+x_2=2\neq 1$), so the resulting state has support outside $\mathcal{F}_0\cap\mathcal{F}_1 = \{010,101\}$.

\section{Constraints with Efficient Gadgets} \label{app: structural encoding}

Circuit details for cardinality equality and inequality follow existing literature while the flow conservation construction is novel to this work and derived in full.

\subsection{Cardinality Equality: \texorpdfstring{$\sum_i x_i = k$}{sum xi = k}}

\paragraph{Feasible set.}

\[
    \mathcal{F}_{\mathrm{eq}}
    = \{x\in\{0,1\}^n \mid \textstyle\sum_{i=0}^{n-1} x_i = k\}.
    \] All feasible strings have Hamming weight $k$, and the number of feasible assignments is $\binom{n}{k}$.

\paragraph{Target state.}

The uniform superposition over all feasible assignments is the
\emph{Dicke state}
\[
    |D^n_k\rangle
    =
    \frac{1}{\sqrt{\binom{n}{k}}}
    \sum_{\substack{x\in\{0,1\}^n \\ |x|=k}}
    |x\rangle .
\] For example, $|D_1^3\rangle = \tfrac{1}{\sqrt{3}}(|100\rangle+|010\rangle+|001\rangle)$.

\paragraph{Circuit construction.}

We use the short-depth Dicke-state circuit of~\cite{bartschi2022short}, which prepares $|D^n_k\rangle$ in depth $O(k\log(n/k))$ without ancilla qubits; we refer the reader to that work for full circuit details.

\paragraph{Mixer.}

Because the feasible subspace consists entirely of weight-$k$ states, the mixer must preserve Hamming weight.
We use the \emph{XY mixer}: \[ U_{\mathrm{XY}}(\beta) =
    \exp\!\left(-i\tfrac{\beta}{2}\sum_{\langle i,j\rangle}(X_iX_j+Y_iY_j)\right),
\] where the sum runs over adjacent pairs in an open or ring topology.
Each $XX+YY$ term exchanges amplitudes between $|01\rangle$ and $|10\rangle$, preserving Hamming weight and mixing all weight-$k$ states.

\subsection{Cardinality Inequality: \texorpdfstring{$\sum_i x_i \le k$}{sum xi <= k}}

\paragraph{Feasible set.}

\[
    \mathcal{F}_{\mathrm{leq}}
    = \{x\in\{0,1\}^n \mid \textstyle\sum_{i=0}^{n-1} x_i \le k\},
    \;
    M(n,k) = \sum_{w=0}^{k} \tbinom{n}{w}.
    \]

\paragraph{Target state.}

The uniform superposition over feasible assignments is \[
    |S_n^{\le k}\rangle
    =
    \frac{1}{\sqrt{M(n,k)}}
    \sum_{w=0}^{k}
    \sqrt{\tbinom{n}{w}}
    \, |D^n_w\rangle .
    \]

\paragraph{Circuit construction.}

The state $|S_n^{\le k}\rangle$ is prepared following~\cite{bartschi2019deterministic,bartschi2022short}.
Specifically, a staircase of $k$ controlled-$RY$ rotations builds the weight superposition $\sum_{w=0}^{k}\alpha_w|1^w 0^{n-w}\rangle$, after which the Dicke-state circuit of~\cite{bartschi2022short} is applied to simultaneously map each weight-$w$ component to $|D^n_w\rangle$.

\paragraph{Mixer.}

The $XY$ mixer preserves a single Hamming weight and therefore cannot mix across different weight sectors.
A \emph{Grover mixer} reflecting around $|S_n^{\le k}\rangle$ is used instead, which mixes across all feasible weights while still confining the state to the feasible subspace.

\subsection{Flow Conservation}\label{app:flow}
\paragraph{Feasible set.}

Consider a network vertex with $p$ incoming edge variables $\mathcal{I} = \{i_0,\ldots,i_{p-1}\}$ and $q$ outgoing edge variables $\mathcal{O} = \{o_0,\ldots,o_{q-1}\}$.
Flow conservation requires equal total flow in and out: \[
    \sum_{\ell\in\mathcal{I}} x_\ell
    =
    \sum_{\ell\in\mathcal{O}} x_\ell ,
\] i.e. $\mathrm{HW}(x_\mathcal{I}) = \mathrm{HW}(x_\mathcal{O})$, where $\mathrm{HW}(x)$ denotes the Hamming weight of $x$.
Writing $W = \min(p,q)$, the number of feasible configurations is \[
    |\mathcal{F}_{\mathrm{flow}}|
    =
    \sum_{w=0}^{W}
    \tbinom{p}{w}\tbinom{q}{w}.
\]

\paragraph{Target state.}

The uniform superposition over all feasible assignments decomposes into matched-weight sectors across the two registers: \[
    |F_{p,q}\rangle
    =
    \frac{1}{\sqrt{|\mathcal{F}_{\mathrm{flow}}|}}
    \sum_{w=0}^{W}
    \sqrt{\tbinom{p}{w}\tbinom{q}{w}}
    \;|D^p_w\rangle \otimes |D^q_w\rangle .
    \] Each weight sector $w$ contributes a product of Dicke states with matching Hamming weight on the two registers.

\paragraph{Circuit construction.}

The state $|F_{p,q}\rangle$ is prepared in four steps, using no ancilla qubits.
Define $\alpha_w = \sqrt{\binom{p}{w}\binom{q}{w}/|\mathcal{F}_{\mathrm{flow}}|}$ for $w=0,\ldots,W$.

\emph{Step~1: Staircase on $\mathcal{I}$.}
Build the superposition $\sum_{w=0}^{W} \alpha_w |1^w 0^{p-w}\rangle_\mathcal{I}$ using a staircase of $W$-many $RY$/$CRY$ rotations on the in-register:
\begin{itemize}
\item $j=0$: $RY(\theta_0)$ on $i_0$, with $\theta_0 = 2\arccos(\alpha_0)$.
\item $j=1,\ldots,W-1$: $CRY(\theta_j)$ on $i_j$, controlled by $i_{j-1}=1$, with
$\theta_j = 2\arccos\!\left(\alpha_j \big/ \sqrt{\textstyle\sum_{s\ge j}\alpha_s^2}\right)$.
\end{itemize}
This leaves $|\mathcal{O}\rangle = |0^q\rangle$ untouched.

\emph{Step~2: CNOT copy to $\mathcal{O}$.}
Entangle the out-register by applying \[
\mathrm{CNOT}(i_j,\; o_j),\quad j = 0,\ldots,W-1.
\] After this step the joint state is $\sum_{w=0}^{W} \alpha_w\, |1^w 0^{p-w}\rangle_\mathcal{I} \otimes |1^w 0^{q-w}\rangle_\mathcal{O}$, where both registers have matching Hamming weights but amplitude concentrated on the states with all 1s packed into the first $w$ positions rather than uniformly spread over all weight-$w$ strings.

\emph{Step~3: Dicke-state preparation on $\mathcal{I}$.}
Reverse the qubit order of $\mathcal{I}$ and apply the Dicke-state circuit of~\cite{bartschi2022short}, which by linearity simultaneously maps each $|0^{p-w}1^w\rangle \to |D^p_w\rangle$ for all $w \le W$, uniformly spreading amplitude within each weight sector of the in-register.

\emph{Step~4: Dicke-state preparation on $\mathcal{O}$.}
Apply the same procedure to $\mathcal{O}$: reverse and prepare, mapping each $|0^{q-w}1^w\rangle \to |D^q_w\rangle$.

Steps~3 and~4 act on disjoint registers and commute.
The final state is \[
\sum_{w=0}^{W} \alpha_w\, |D^p_w\rangle \otimes |D^q_w\rangle
= |F_{p,q}\rangle.
\]

\begin{figure}[t]
\centering
%
\begin{tikzpicture}[
    Sgate/.style={shading=axis, left color=rpBlue!65,       right color=rpBlue!25,       shading angle=135},
    IDgate/.style={shading=axis, left color=rpGreen!65,     right color=rpGreen!25,      shading angle=135},
    ODgate/.style={shading=axis, left color=rpLightBlue!65, right color=rpLightBlue!25,  shading angle=135},
]
\node{%
\begin{quantikz}[row sep=1.0em, column sep=0.8em]
  \lstick{$i_0$}
    & \gate[style={Sgate}]{RY(\theta_0)}
      \gategroup[4, steps=2,
        style={inner sep=-2pt, dashed, rounded corners},
        label style={}]
        {\small Step 1}
    & \ctrl{1}
    & \ctrl{2}
      \gategroup[4, steps=2,
        style={inner sep=-2pt, dashed, rounded corners},
        label style={}]
        {\small Step 2}
    & \qw
    & \gate[2, swap, style={draw=none, fill=none}]{}
      \gategroup[4, steps=2,
        style={inner sep=-2pt, dashed, rounded corners},
        label style={}]
        {\small Steps 3\,\&\,4}
    & \gate[2, style={IDgate}]
        {\substack{|D^p_w\rangle\\\text{prep}}}
    & \qw \\
  \lstick{$i_1$}
    & \qw
    & \gate[style={Sgate}]{RY(\theta_1)}
    & \qw
    & \ctrl{2}
    & \qw
    & \qw
    & \qw \\
  \lstick{$o_0$}
    & \qw
    & \qw
    & \targ{}
    & \qw
    & \gate[2, swap, style={draw=none, fill=none}]{}
    & \gate[2, style={ODgate}]
        {\substack{|D^q_w\rangle\\\text{prep}}}
    & \qw \\
  \lstick{$o_1$}
    & \qw
    & \qw
    & \qw
    & \targ{}
    & \qw
    & \qw
    & \qw
\end{quantikz}
};
\end{tikzpicture}
\caption{State-preparation circuit for the flow conservation constraint ($p = q = 2$, $W = \min(p,q) = 2$). Step~1: a staircase of $RY$/$CRY$ rotations on $\mathcal{I}$ builds the weight superposition ($RY$ for $j=0$, $CRY$ controlled by $i_{j-1}$ for $j \geq 1$). Step~2: CNOT gates copy the weight to $\mathcal{O}$. Steps~3--4: SWAP gates reverse qubit order within each register, after which $U_{\mathrm{Dicke}}$~\cite{bartschi2022short} maps each weight-$w$ component to $|D^p_w\rangle$ (resp.\ $|D^q_w\rangle$). Steps~3 and~4 act on disjoint registers and commute. Dashed boxes show the joint state produced after each step.}
\label{fig:flow_circuit}
\end{figure}

\paragraph{Concrete example ($p = q = 2$).}

With two incoming and two outgoing wires the weight sectors are $w\in\{0,1,2\}$ with $|\mathcal{F}| = 1+4+1 = 6$ and $\alpha_0=\alpha_2 = 1/\sqrt{6}$, $\alpha_1 = 2/\sqrt{6}$.
\begin{enumerate}
    \item \emph{Staircase on $i_0, i_1$}:
      $RY\!\left(2\arccos\!\tfrac{1}{\sqrt{6}}\right)$ on $i_0$; $CRY(\theta_1)$ on $i_1$ controlled by $i_0$, where $\theta_1 = 2\arccos\!\sqrt{4/5}$.
    \item \emph{CNOT copy}: $\mathrm{CNOT}(i_0, o_0)$;
      $\mathrm{CNOT}(i_1, o_1)$.
    \item \emph{Dicke-state preparation on $i_0, i_1$}: the circuit of~\cite{bartschi2022short} maps
      $|11\rangle\to|11\rangle$, $|10\rangle\to\tfrac{1}{\sqrt{2}}(|10\rangle+|01\rangle)$, $|00\rangle\to|00\rangle$.
    \item \emph{Dicke-state preparation on $o_0, o_1$}: same procedure applied to the out-register.
\end{enumerate}
The result is $\tfrac{1}{\sqrt{6}}\bigl(|00\rangle|00\rangle + |10\rangle|10\rangle + |10\rangle|01\rangle + |01\rangle|10\rangle + |01\rangle|01\rangle + |11\rangle|11\rangle\bigr)$.

\paragraph{Mixer.}

Applying an independent Ring-XY mixer on $\mathcal{I}$ and on $\mathcal{O}$ preserves the Hamming weight within each register separately, hence preserves $\mathrm{HW}(\mathcal{I})=\mathrm{HW}(\mathcal{O})$.
However, this does \emph{not} connect the full feasible set: the feasible space decomposes as \[
\mathcal{H}_{\mathrm{flow}}
=
\bigoplus_{w=0}^{W}
\mathrm{span}\!\left\{
|x_\mathcal{I}\rangle |x_\mathcal{O}\rangle :
\mathrm{HW}(x_\mathcal{I})=\mathrm{HW}(x_\mathcal{O})=w
\right\},
\] and independent Ring-XY mixers act only within each fixed-$w$ sector.
Thus they cannot move amplitude between sectors with different common weight.
To mix over the entire feasible subspace while preserving feasibility, we instead use the Grover mixer built from the prepared feasible superposition: \[ U_B(\beta) = e^{-i\beta |F_{p,q}\rangle\langle F_{p,q}|}.
\] This is the mixer used for the flow-conservation constraint in the main text.

\section{Variational Constraint Gadget Training Details and Results}
\label{app:vcg}

\subsection{Preprocessing and Exact Reductions}

Before variational training, we inspect the structure of each constraint to identify cases admitting exact preparation.
If the feasible set $\mathcal{F}_k$ contains a single assignment, the corresponding state is prepared exactly using $X$ gates.
If $\mathcal{F}_k$ corresponds to a union of Hamming-weight classes (e.g., a cardinality inequality), the target state can be prepared as a superposition of Dicke states without training.

Of the 60 constraint instances considered (30 knapsack, 30 quadratic knapsack), 14 admit such exact constructions and are assigned depth $p=0$.

\subsection{Training Procedure and Hyperparameters}

We use the term \emph{training} to refer to the process of optimizing circuit parameters via gradient descent.
For constraints requiring variational construction, we train parameterized circuits using a two-stage procedure:

\begin{enumerate}
    \item \textbf{Warm start:} A single-layer QAOA circuit is optimized over 5 random initializations (150 steps each) to obtain a good starting point.
    \item \textbf{Layer-wise growth:} Additional layers are added sequentially (ma-QAOA), initializing each new layer from the previous optimum and refining via 20 restarts of 200 steps each.
\end{enumerate}

Training uses the Adam optimizer (learning rate $0.05$) and halts when both $P_{\mathcal{F}_k} \geq 0.999$ and $\mathcal{S}_{\mathrm{norm}} \geq 0.9999$ are satisfied, or when a maximum of $p_{\max} = 8$ layers is reached.
Remaining hyperparameters are listed in Table~\ref{tab:vcg_hyperparams}.

\begin{table}[h]
\centering
\caption{VCG training hyperparameters.}
\label{tab:vcg_hyperparams}
\begin{tabular}{lc}
\hline
\textbf{Parameter} & \textbf{Value} \\
\hline
QAOA warm-up restarts & 5 \\
QAOA warm-up steps    & 150 \\
ma-QAOA restarts per layer & 20 \\
ma-QAOA steps per restart  & 200 \\
Learning rate (Adam)       & 0.05 \\
Feasible probability threshold $P_{\mathcal{F}_k}$ & $0.999$ \\
Entropy threshold $\mathcal{S}_{\mathrm{norm}}$ & $0.9999$ \\
Maximum layers ($p_{\max}$) & 8 \\
Measurement shots          & 10,000 \\
\hline
\end{tabular}
\end{table}

\end{document}